\documentclass[conference]{IEEEtran}
\usepackage{caption}
\IEEEoverridecommandlockouts

\usepackage[table]{xcolor}
\usepackage{etoolbox}
\usepackage{comment}
\usepackage{pdfcomment}
\usepackage{subcaption}
\usepackage{amsmath}

\usepackage{biblatex} 
\definecolor{dark-red}{rgb}{0.8, 0.0, 0.0}
\definecolor{columbiablue}{rgb}{0.21, 0.27, 0.7}
\definecolor{columbiablue2}{rgb}{0.18, 0.24, 0.69}
\definecolor{columbiablue3}{rgb}{0.20, 0.26, 0.70}
\definecolor{columbiaazure3}{rgb}{0.10, 0.42, 0.72}
\definecolor{orobrillante}{rgb}{1.00, 0.84, 0.00}
\definecolor{orometallico}{rgb}{0.80, 0.55, 0.07}
\definecolor{darkcyan}{rgb}{0.0, 0.0, 0.8}
\definecolor{electricviolet}{rgb}{0.56, 0.0, 1.0}
\definecolor{forestgreen}{rgb}{0.0, 0.5, 0.0}
\definecolor{darkorange2}{rgb}{0.95, 0.35, 0.00}
\definecolor{brown}{rgb}{0.55, 0.25, 0.0}
\definecolor{blue-violet}{rgb}{0.54, 0.17, 0.89}
\definecolor{dartmouthgreen}{rgb}{0.05, 0.5, 0.0}
\definecolor{light-blue}{rgb}{0.5, 0.5, 0.99}
\definecolor{brinkpink}{rgb}{0.85, 0.25, 0.5}
\definecolor{cyan(process)}{rgb}{0.0, 0.55, 0.85}
\definecolor{deeplilac}{rgb}{0.6, 0.33, 0.73}
\definecolor{indaco}{rgb}{0.25, 0.0, 1.0}
\definecolor{indaco2}{rgb}{0.35, 0.18, 0.98}
\definecolor{verylightgray}{rgb}{0.8,0.8,0.8}
\definecolor{verylightgray}{rgb}{0.8,0.8,0.8}

\newcommand{\MB}[1]{\textcolor{dark-red}{{\bf MB:}{\small \texttt{~#1~}}}}
\newcommand{\FC}[1]{\textcolor{columbiaazure3}{{\bf FC:}{\small \texttt{~#1~}}}}

\newcommand{\DV}[1]{\textcolor{orometallico}{{\bf DV:}{\small \texttt{~#1~}}}}
\newcommand{\dv}[1]{\textcolor{orometallico}{#1}}
\newcommand{\FV}[1]{\textcolor{darkorange2}{{\bf FV:}{\small \texttt{~#1~}}}}

\usepackage{orcidlink}
\usepackage[a4paper,top=1.75cm,bottom=1.75cm,left=1.5cm,right=1.5cm]{geometry}
\usepackage{amsmath,amssymb,mathtools}
\usepackage{booktabs}
\usepackage{enumitem}
\usepackage{algorithm}
\usepackage[noend]{algpseudocode}
\usepackage{microtype}
\usepackage{hyperref}
\usepackage[dvipsnames]{xcolor}
\hypersetup{unicode=true, colorlinks=true,linkcolor=blue,urlcolor=blue,citecolor=blue}

\newcommand{\E}{\mathbb{E}}
\newcommand{\Var}{\mathbb{V}\mathrm{ar}}

\newcommand{\RR}{\mathbb{R}}

\newcommand{\ptf}[1]{\mathcal{#1}}
\newcommand{\initp}{\ptf{P}}
\newcommand{\bp}{\ptf{B}}
\newcommand{\p}[1]{\mathbf{#1}}
\newcommand{\m}[1]{\mathbf{#1}}

\newcommand{\pls}[2]{z_{#1}\left(#2\right)}
\newcommand{\pl}[1]{\mathbf{z}\left(#1\right)}
\newcommand{\pT}{\ptf{T}}

\newcommand{\pTx}[1]{\ptf{T}\left(#1\right)}
\newcommand{\Cov}{\mathrm{Cov}}
\newcommand{\obj}{\mathrm{obj}}
\newcommand{\objM}{\mathrm{obj}_{\mathrm{M}}}
\newcommand{\Vega}{\mathcal{V}}

\DeclareMathOperator*{\argmin}{arg\,min}

\newcommand{\Gurobi}{\textsc{Gurobi}}

\newcommand{\knds}{\kern-\nulldelimiterspace}

\newcommand{\nU}{N_{\text {UEI}}}
\newcommand{\nN}{N_{\text{notional}}}
\newcommand{\nS}{S}
\definecolor{verylightgray}{rgb}{0.8,0.8,0.8}
\definecolor{verylightgray}{rgb}{0.8,0.8,0.8}
\definecolor{brown}{rgb}{0.55, 0.25, 0.0}
\definecolor{forestgreen}{rgb}{0.0, 0.5, 0.0}
\definecolor{blue-violet}{rgb}{0.54, 0.17, 0.89}
\definecolor{dartmouthgreen}{rgb}{0.05, 0.5, 0.0}
\definecolor{dark-red}{rgb}{0.8, 0.0, 0.0}
\definecolor{light-blue}{rgb}{0.5, 0.5, 0.99}
\definecolor{brinkpink}{rgb}{0.85, 0.25, 0.5}
\definecolor{columbiablue}{rgb}{0.61, 0.87, 1.0}
\definecolor{cyan(process)}{rgb}{0.0, 0.55, 0.85}
\definecolor{darkcyan}{rgb}{0.0, 0.0, 0.8}
\definecolor{darkorange}{rgb}{0.45, 0.20, 0.0}
\definecolor{darkviolet}{rgb}{0.19, 0.10, 0.20}		
\definecolor{deeplilac}{rgb}{0.6, 0.33, 0.73}
\definecolor{electricultramarine}{rgb}{0.25, 0.0, 1.0}
\definecolor{electricviolet}{rgb}{0.56, 0.0, 1.0}

\newcommand{\parab}[1]{\noindent\textbf{#1}}

\title{Quantum-Inspired Risk Portfolio Optimization\\ via Exponential Tilting}
\author{
    \IEEEauthorblockN{
    Fabio Vitale\IEEEauthorrefmark{2}\orcidlink{0009-0000-7398-9436},
    Francesca Cibrario\IEEEauthorrefmark{1}\orcidlink{0009-0007-8290-4992},
    Davide Veronelli\IEEEauthorrefmark{1}\orcidlink{0000-0002-3810-6287},
    Chiara Vercellino\IEEEauthorrefmark{3}\orcidlink{0000-0002-0562-3157},
    Valeria Zaffaroni\IEEEauthorrefmark{1}\orcidlink{0009-0002-8221-5904},\\
    Giacomo Vitali\IEEEauthorrefmark{3}\orcidlink{0000-0002-3056-796X},
    Gregorio Antonio Polito\IEEEauthorrefmark{1}\orcidlink{0009-0001-8532-1713},
    Emanuele Dri\IEEEauthorrefmark{3}\orcidlink{0000-0002-5144-1514},\\
    Carla Monferrato\IEEEauthorrefmark{2}\orcidlink{0009-0007-2140-8186},
    Olivier Terzo\IEEEauthorrefmark{3}\orcidlink{0000-0001-8482-2607},
    Davide Corbelletto\IEEEauthorrefmark{1}\orcidlink{0009-0003-8830-2619},
    Marco Bianchetti\IEEEauthorrefmark{1}\orcidlink{0000-0003-3750-087X}
    }
    \\
    \IEEEauthorblockA{
    \IEEEauthorrefmark{1}Intesa Sanpaolo, Torino, Italy\\
    francesca.cibrario@intesasanpaolo.com, 
    davide.veronelli@intesasanpaolo.com,\\
    valeria.zaffaroni@intesasanpaolo.com,
    gregorio.polito@intesasanpaolo.com,\\
    davide.corbelletto@intesasanpaolo.com,
    marco.bianchetti@intesasanpaolo.com,\vspace{1.5mm}\\
    \IEEEauthorrefmark{2}Intesa Sanpaolo Innovation Center, Torino, Italy\\
    fabio.vitale@intesasanpaolo.com,
    carla.monferrato@intesasanpaolo.com,\vspace{1.5mm}\\
    \IEEEauthorrefmark{3}Fondazione LINKS, Torino, Italy\\
    chiara.vercellino@linksfoundation.com,
    giacomo.vitali@linksfoundation.com,\\
    emanuele.dri@linksfoundation.com,
    olivier.terzo@linksfoundation.com
    }
}

\date{\today}

\def\BibTeX{{\rm B\kern-.05em{\sc i\kern-.025em b}\kern-.08em
    T\kern-.1667em\lower.7ex\hbox{E}\kern-.125emX}}

\hypersetup{bookmarks=false}
\begin{document}
\maketitle
\begin{abstract}
Financial portfolio optimization is often framed through mean--variance objectives, yet real profit-and-loss (P\&L) scenario distributions are typically discrete and non-Gaussian. 
To explore realistic P\&L distributions, we propose an objective function based on exponential tilting and certainty equivalent under constant absolute risk aversion. We then derive a second-order approximation, recovering a quadratic optimization problem solvable through mixed integer quadratic programming (MIQP) and quadratic unconstrained binary optimization (QUBO).
Constraints in the QUBO formulation are incorporated via Lagrangian relaxation and slack variable approaches, enabling a direct comparison between classical constrained MIQP and quantum-inspired methods.
In order to study how exponential tilting affects mean, standard deviation, and tail-risk measures such as Value at Risk and Conditional Value at Risk, we design an experimental protocol based on scanning a grid of tilt parameters and solving the associated optimization problems.
The results on a realistic trading portfolio show how the tilts can be tuned to optimize different risk-return profiles while remaining competitive with those found using other state-of-the-art approaches.
Moreover, performance analysis suggests that, even using classical solvers, the QUBO formulation with Lagrangian relaxation scales better than the MIQP approach, providing the best constraint encoding strategy.
\end{abstract}


\section{Introduction}

Portfolio optimization is the process of selecting the best asset allocation
among available choices in order to balance expected return and risk.
In a scenario-based setting, portfolio performance is evaluated over a finite collection of possible market states, 
each producing a profit-and-loss (P\&L) outcome, i.e., a change in the economic value of the portfolio.
In this framework, a standard approach in quantitative finance is the {\em mean--variance} criterion, introduced by Markowitz \cite{e5a1bb8f-41b7-35c6-95cd-8b366d3e99bc}, which selects portfolios by maximizing the average P\&L across scenarios (the mean) while penalizing the variability of P\&L across scenarios (the variance), thus favoring portfolios that achieve higher returns
with limited dispersion of outcomes.
This approach is frequently applied to relatively simple investment portfolios (e.g., $N$ stocks with continuous quantities) and is suitable when the P\&L distribution is Gaussian, since in that case the portfolio risk is fully characterized by the variance.
However, heterogeneous trading portfolios of commercial or investment banks typically encompass millions of financial instruments (such as stocks, securities, loans, derivatives, funds, etc.), featuring any level of complexity (plain-vanilla, exotics, etc.), depending on a wide range of underlying risk factors (interest rates, equity, etc.). The resulting P\&L distributions can be markedly non-Gaussian, implying that risk cannot be fully captured by variance alone. Practitioners may therefore wish to look at the P\&L distribution from different perspectives, for example by focusing on adverse or favorable outcomes.
Moreover, in this realistic setting, the portfolio optimization problem consists in defining a strategy to build a new portfolio $\pT$ improving an already owned \emph{initial portfolio} $\initp$, adding an appropriate selection of liquid financial instruments.

The objective of this work is to deal with this configuration and go beyond the mean–-variance criterion while preserving a quadratic objective function. This enables the use of mixed-integer quadratic programming to solve small- and medium-scale portfolio optimization problems with current classical computing resources, thereby paving the way for the application of quantum computing to larger instances.

Our approach starts from the initial portfolio $\initp$, identifies a suitable portfolio $\bp$, hereafter called \emph{base-point} portfolio, and uses a softmax map applied to its P\&L distribution to reweight scenarios and obtain a new non-uniform probability measure. This construction is equivalent to the \emph{Esscher transform} or exponential tilting~\cite{gerber1994option,papapantoleon2008introduction}.
The portfolio $\bp$ is ``suitable" in the sense that it should be easily computable and reasonably in a neighborhood of promising solutions.
A tilt parameter controls how P\&L scenarios are reweighted: positive tilts concentrate weight on scenarios corresponding to high P\&L values of $\bp$, and conversely for negative tilts. 

In order to find an optimal portfolio $\pT$, we maximize a single monetary value, the \emph{certainty equivalent} (CE) \cite{gollier2001economics}, associated with a P\&L distribution under the weighted probability measure.
To define the CE, we use the constant absolute risk aversion (\emph{CARA}) exponential utility function~\cite{gollier2001economics,gerber1998utility}, which provides a tractable way of penalizing risk while maintaining a clear monetary interpretation.
Finally, we replace the exact CE with its second-order approximation, which can be used as a tilted mean--variance quadratic objective function.

To validate the proposed theoretical framework, we adopt the setup introduced in \cite{bianchetti2025riskawaretradingportfoliooptimization}, where the initial trading portfolio is optimized by selecting a specific set (and related discretized quantities) of \emph{unique eligible instruments} (UEIs) subject to predefined constraints. 
The base-point portfolio required by our approach is taken to be the solution of a constrained mean-–variance relaxed problem using quadratic programming (QP), with continuous decision variables representing the quantity of each UEI added to the initial portfolio. 
For a grid of tilt values, we construct the corresponding objective functions and solve the constrained discretized problem using MIQP to evaluate the impact of tilting. Finally, for a subset of selected tilts, we demonstrate that the proposed approach can, in principle, be addressed using quantum computing techniques such as quantum annealing or variational quantum algorithms. We formulate a QUBO model by embedding constraints into penalty terms,  using both slack variables and a Lagrangian relaxation approach. The QUBO formulation is validated on classical hardware using the solver \Gurobi~\cite{gurobi2026} and the quantum-inspired Simulated Bifurcation method \cite{Ageron_Simulated_Bifurcation_SB_2023}, demonstrating the superior performance of the Lagrangian relaxation-based constraint encoding strategy. Notably, a preliminary scaling analysis, conducted by increasing the size of the optimization problem, demonstrates that on classical hardware the unconstrained approach can provide feasible results, requiring fewer hardware resources than the constrained version, while the quantum-inspired methodology demonstrates better memory management.

\smallskip
\parab{Main contributions.}
\begin{itemize}
\item \textbf{Exponential tilting as a weighting method.}
We introduce a one-parameter family of Esscher/softmax scenario weights around a fixed base-point portfolio of interest. Negative tilts emphasize downside base-point scenarios, while positive tilts emphasize upside ones. 

\item \textbf{Quadratic objective function with clear financial interpretation.}
Using a CARA certainty equivalent, approximated to the second-order, we obtain a quadratic objective function with a clear risk--return meaning: it rewards expected gains and penalizes residual variability under the tilted scenario measure, while retaining non-Gaussian distributional information through the tilted weights.

\item \textbf{QUBO formulation and constraints encoding.}
The quadratic objective function is suitable for a QUBO formulation and, therefore, for quantum optimization.  We embed financially relevant constraints into the quadratic objective function through penalties and Lagrangian relaxation, avoiding the introduction of additional binary variables.

\item \textbf{Experimental validation and scalability.} 
Finally, experimental results validate the practical effectiveness of the proposed formulation, showing that the QUBO formulation with Lagrangian relaxation achieves competitive performance while preserving computational tractability on classical hardware.

\end{itemize}


\section{Related Work}

\subsection{Risk-aware portfolio optimization}
\label{sec:rw-riskaware}

Recent work in risk-aware trading portfolio optimization is formulated as the selection of a \emph{tradable} strategy that improves a given objective function of an initial portfolio while satisfying risk and business constraints. A representative framework is Risk-aware Trading Portfolio Optimization (RATPO) and its particle-swarm solver RATS~\cite{bianchetti2025riskawaretradingportfoliooptimization}.

In RATPO, the starting point is an initial portfolio 
(e.g., related to a single trading desk or the whole trading book) at some fixed date, and a user-specified universe of eligible optimization instruments, i.e., the liquid market instruments quoted by exchanges, brokers, or market makers, typically used by traders to hedge Delta, Vega, and Gamma risks.
RATS introduces the notion of unique eligible instruments: each UEI is an eligible optimization instrument identified by a tuple of discrete parameters (e.g., underlying, maturity, strike, option type) with unitary notional value. 
An \emph{eligible optimization strategy} (EOS) is then a combination of UEIs weighted with their corresponding discrete notional amounts, subject to constraints that encode feasibility and risk limits.
Operationally, the decision variables specify which UEI-notional value combinations are selected, i.e., elements of a discrete Cartesian product of instrument-parameters and notional value domains~\cite{bianchetti2025riskawaretradingportfoliooptimization},
rather than continuous portfolio weights in a frictionless mean--variance model.
The objective function includes market risk measures commonly used for risk management and capital purposes, such as Value at Risk
(VaR) and Conditional Value at Risk (CVaR), calculated from empirical discrete profit-and-loss (P\&L) distributions obtained by full portfolio revaluation in historical risk-factor scenarios.
VaR is a market risk metric that estimates the maximum potential loss of a portfolio over a defined period within a specific confidence level. CVaR, a.k.a. Expected Shortfall (ES), measures the average loss exceeding the VaR threshold. It is often used to provide a more robust measure of extreme loss than VaR alone~\cite{puelz2001var,rockafellar2000cvar}.
The constraints addressed in RATPO are the following.
\begin{itemize}
    \item \textbf{EOS composition}: prescribes the EOS composition in terms of maximum number and types of UEIs (e.g., one Futures or single stock and two call/put options) for each relevant underlying risk factor, encoding the traders' optimization strategy.
    \item \textbf{UEI quantities}: prescribes discrete notional values from a fixed discrete set for each UEI, encoding the market tradable sizes.
    \item \textbf{Greeks constraints}: prescribes the maximum allowed variation of Delta, Gamma and Vega sensitivities, encoding the rigidity of the portfolio risk profile decided by traders.
\end{itemize}

Our work is based on constraints, discrete instrument choices, and datasets (see Section~\ref{subsec:datasets} for more details) introduced in RATPO. 
It complements classical mean--variance and VaR/CVaR-based portfolio optimization~\cite{rockafellar2000cvar,wozabal2010dcvar,soleimani2009ga,puelz2001var}, building a novel quadratic objective function without imposing convexity or differentiability, as for particle-swarm optimization~\cite{shi1998pso,bianchetti2025riskawaretradingportfoliooptimization}. In contrast with meta-heuristic search over structured trade sets, our approach keeps a single MIQP/QUBO solve by building a quadratic surrogate anchored at a fixed base-point, while still allowing tail emphasis through tilts and scenario reweighting. To build our objective function, we use an explicit Esscher/softmax reweighting of scenarios based on base-point P\&L scores. 
Esscher transforms are classical in quantitative finance, for example, in option pricing and in L\'evy-based asset models~\cite{gerber1994option,papapantoleon2008introduction}. Related ideas of scenario reweighting also appear in entropy-pooling approaches, where posterior scenario probabilities are obtained by relative-entropy minimization under stress views~\cite{meucci2008entropypooling}.

\subsection{Quantum portfolio optimization}

Combinatorial optimization consists of finding the optimum solution of a problem defined by an objective function from a finite set of candidates. A brute-force strategy that evaluates all possible candidate solutions becomes computationally intractable as the dimension of the problem increases. Classical heuristics can be used to approach large scale combinatorial optimization problems providing high-quality solutions within reasonable computational time, without guaranteeing solution optimality. In this context, quantum optimization has emerged as an alternative computational paradigm that aims to tackle combinatorial problems through quantum computing.

Quantum computation can be realized through two principal paradigms: gate-based and analog quantum computing.
QUBO models, which can be directly mapped to Ising Hamiltonians, provide a common mathematical framework compatible with both approaches. Gate-based quantum computers allow for the solution of QUBO problems through hybrid quantum–classical algorithms, such as the quantum approximate optimization algorithm (QAOA) \cite{farhi2014quantumapproximateoptimizationalgorithm}, in which a parameterized quantum circuit is iteratively updated through a classical loop minimizing the expectation value of the Hamiltonian.
Alternatively, analog quantum approaches, such as quantum annealing \cite{FINNILA1994343}, map the Hamiltonian into the system's energy and exploit adiabatic evolution to drive the system toward its ground state, i.e. the optimum of the optimization problem.

A key challenge in applying these paradigms to industrial combinatorial optimization problems, which typically involve constraints, continuous or non-binary variables, and non-quadratic objective functions, lies in translating such problems into QUBO form.

A comprehensive review of quantum approaches applied to portfolio optimization is provided in \cite{Naik_Yeniaras_Hellstern_Prasad_Vishwakarma_2025}. From early studies based on mean--variance portfolio optimization formulated as QUBO problems~\cite{Marzec2013}, subsequent research has progressed in two main directions: (i) QUBO reformulations of more realistic portfolio models and (ii) hybrid quantum–classical approaches to extend the perimeter of addressable models relaxing QUBO requirements. 

Preserving the QUBO formulation enables hardware-agnostic implementations by encoding non-quadratic or constrained portfolio optimization problems into the quadratic binary objective while remaining faithful to the original financial intent. Linear inequality constraints are commonly handled by introducing slack variables \cite{qubo} or by unbalanced penalization \cite{Monta_ez_Barrera_2024}. The former approach improves constraint enforcement, but increases the number of binary variables, thus expanding the effective search space and the associated quantum resource requirements. The latter, while variable efficient, requires careful coefficient tuning to avoid infeasible low energy states. Higher-order objective functions and nonlinear constraints can be reduced to quadratic form by introducing auxiliary variables \cite{mandal2020compressedquadratizationhigherorder}. 
Following this direction, several approaches have been proposed beyond the standard mean--variance QUBO approach. In \cite{PhillipsonBhatia2021} portfolio variance minimization is combined with a linear penalty enforcing a minimum return threshold instead of directly minimizing it. Binary variables indicate whether a specific asset is selected. In \cite{math12091291} and \cite{quantum6020018}, binary variables instead encode discretized asset allocation quantities. The former proposes QUBO formulations based on the Herfindahl–Hirschman Index as a risk measure and return on capital as a performance metric. The latter instead proposes a QUBO objective derived from the Sharpe ratio.   

The second direction of research takes advantage of the flexibility of hybrid quantum–classical algorithms to relax the requirement of quadratic unconstrained objective functions. These methods allow the inclusion of higher-order or nonlinear objectives and explicit constraints management 
\cite{a15060202}, but often require deeper quantum circuits or non-native interactions, which may limit their applicability on current noisy intermediate-scale quantum (NISQ) hardware.   
In \cite{Uotila_Ripatti_Zhao_2025}, higher-order portfolio optimization is addressed using gate-based QAOA, extending the Hamiltonian to incorporate skewness and kurtosis. Additionally, Value-at-Risk (VaR) objectives have been incorporated into the optimization process by combining quantum annealing with an external classical optimization loop \cite{Xu_Dasgupta_Pothen_Banerjee_2023}.

This work proposes a QUBO preserving reformulation inspired by the portfolio optimization framework described in~\cite{bianchetti2025riskawaretradingportfoliooptimization}, designed to remain compatible with both gate-based and analog quantum paradigms. By limiting the introduction of auxiliary decision variables, the approach aims to facilitate future implementation on larger-scale quantum hardware.



\section{Scenario-based framework and tilting}
\label{sec:tilting}

This section introduces the scenario-based framework and the exponential tilting used throughout the paper. We describe (i) the problem setting, introducing P\&Ls and the decision variable, (ii) the relevant metrics used in the optimization process (tilted P\&L mean and variance under Esscher/softmax weights), (iii) the Markowitz optimization setting and the choice of the tilting base-point $\bar{\mathbf{x}}$.

\subsection{Problem setting}\label{subsec:pl_decomposition}
Let $\nU$ be the number of possible UEIs, each described by its specific tuple of discrete parameters that, with the corresponding notional values, make up the EOS.
For each UEI, we choose discrete notional values from its corresponding domain with cardinality equal to $\nN$ (see Section~\ref{sec:rw-riskaware}).
Thus, we may enumerate all possible pairs UEI-notional value using an index $u = 1 ,\ldots,d$ with $d:=\nU\cdot \nN$.

Let $s = 1,\ldots,\nS$ index the risk factors scenarios and the corresponding P\&L scenarios. A P\&L scenario represents the change of the economic value of the portfolio according to the risk factors scenario. 
Then, for each $u$ and $s$, let $X_{s,u}$ denote the
\emph{unit-notional} P\&L contribution of UEI indexed by $u$ in scenario $s$, and let $\theta_u$ denote its notional value. The matrix $\m{G}\in\mathbb{R}^{\nS\times d}$ collects the notional value-scaled scenario contributions $G_{s,u} := \theta_u\,X_{s,u}$.
The P\&L distribution of any portfolio $\ptf{H}$ is indicated by the vector $\pl{\ptf{H}}\in\RR^\nS$ whose $s$-th component is $\pls{s}{\ptf{H}}$.

In our setting, an EOS
can be identified by a decision vector $\p{x}\in\{0,1\}^d$, whose $u$-th component takes the value $1$ if the pair UEI-notional value indexed by $u$ is to be added to the initial portfolio $\initp$ and $0$ otherwise.
Every EOS $\p{x}$ defines a new portfolio $\pTx{\p{x}}$, obtained by adding the EOS to $\initp$, whose P\&L distribution is
$$
\pl{\pTx{\p{x}}} = \pl{\initp} + \m{G}\p{x}.
$$

\subsection{Weighted metrics and exponential tilting}\label{subsec:stressed-moments}
A probability measure can be identified with a real vector $\mathbf{w}=(w_1,\dots,w_\nS)^\top$, whose components are non-negative, sum to $1$ and represent the probability of each scenario.
Given two vectors $\mathbf{f},\mathbf{g}\in\RR^\nS$, representing the values of functions defined on the scenario set, we can compute their expected value and covariance under $\mathbf{w}$
\begin{align}
    \E_\mathbf{w}[\mathbf{f}] & = \sum_{s=1}^\nS w_sf_s = \mathbf{f}^\top\mathbf{w},\label{eq:w_expected_value}\\
    \Cov_\mathbf{w}(\mathbf{f},\mathbf{g}) & = \sum_{s=1}^\nS w_sf_sg_s - \sum_{s,s'=1}^\nS w_sw_{s'}f_sg_{s'}\nonumber\\
    & = \mathbf{g}\!^\top\!\left(\mathrm{diag}(\mathbf{w}) - \mathbf{w} \mathbf{w}^\top\right) \mathbf{f}  \label{eq:w_cov}\,.
\end{align}

Observe that Eq. \eqref{eq:w_expected_value} and \eqref{eq:w_cov} 
can be extended to matrices 
 $\m{M}\in\mathbb{R}^{\nS\times d}$, so that 
 $$
 \begin{aligned}
\E_\mathbf{w}[\m{M}] & = \m{M}^\top\m{w} \in\mathbb{R}^{d}\,, \\
\Cov_\mathbf{w}(\m{f},\m{M}) & =\mathbf{M}^\top\!\left(\mathrm{diag}(\mathbf{w}) - \mathbf{w} \mathbf{w}^\top\right) \m{f} \in\mathbb{R}^{d}\,,\\
\Cov_\mathbf{w}(\m{M},\m{M}) & =\mathbf{M}^\top\!\left(\mathrm{diag}(\mathbf{w}) - \mathbf{w} \mathbf{w}^\top\right) \m{M} \in\mathbb{R}^{d\times d}\,.
\end{aligned}
$$

Assume that the base-point portfolio $\bp$ is $\pTx{\bar{\p{x}}}$ for a given $\bar{\p{x}}$ (defined in Section \ref{subsec:mark_and_basepoint}).
The exponential tilting is defined by reweighting scenarios through an exponential score. For a scalar tilt parameter $t\in\mathbb{R}$, we define the exponentially tilted (Esscher) weights vector $\boldsymbol{\pi}(t|\bp)$, having as $s$ entry

\begin{equation}\label{eq:esscher-weights}
\pi_s(t|\bp) :=
\frac{w_s\,\exp\bigl(t\, \pls{s}{\bp}\bigr)}
{\sum_{s'=1}^\nS w_{s'}\,\exp\bigl(t\,  \pls{s'}{\bp}\bigr)}\,.
\end{equation}
In a finite scenario set, Eq. \eqref{eq:esscher-weights} is a softmax reweighting of the base-point P\&L $\pl{\bp}$: as $|t|$ increases, the weights become more concentrated in extreme scenarios, while $t=0$ recovers the base weights $\mathbf{w}$.
For $t > 0$ the reweighting emphasizes positive P\&L scenarios, whereas for $t < 0$ it emphasizes negative P\&L scenarios.

We define the tilted mean vector $\bar{\mathbf{b}}_{t,\bp}\in\mathbb{R}^d$ and the covariance matrix of scenario contributions $\bar{\mathbf{Q}}_{t,\bp}\in\mathbb{R}^{d\times d}$ as the exponentially weighted sample mean and covariance
\[
\bar{\mathbf{b}}_{t,\bp} := \mathbb{E}_{\boldsymbol{\pi}(t|\bp)}[\m{G}],
\qquad
\bar{\mathbf{Q}}_{t,\bp} := \mathrm{Cov}_{\boldsymbol{\pi}(t|\bp)}(\m{G},\m{G}).
\]
Additional details are provided in Appendix \ref{app:setup-basic-identities}. 

We can define the P\&L \emph{difference} with respect to the base-point 
\begin{equation}\label{eq:inc-pnl}
\mathbf{y}(\mathbf{x}) := \pl{\pTx{\mathbf{x}}}-\pl{\pTx{\bar{\mathbf{x}}}}
= \mathbf{G}\bigl(\mathbf{x}-\bar{\mathbf{x}}\bigr).
\end{equation}

We work with $\mathbf{y}(\mathbf{x})$
for three reasons:
(i)~the static term~$ \pl{\initp}$ cancels out, so the tilted moments
depend only on the decision-dependent part;
(ii)~under CARA preferences, translation invariance makes the
comparison between portfolios naturally depending on 
P\&L difference relative to a fixed base-point, so the absolute level
$\pl{\pTx{\bar{\mathbf{x}}}}$ does not affect the solution of the exact optimization problem (see Section \ref{sec:objective} for details on CARA);
(iii)~with $\bar{\mathbf{x}}$ chosen as a candidate in the neighborhood of good solutions,
deviations from $\bar{\mathbf{x}}$ under the tilted surrogate
reflect the effect of exponential reweighting rather than correction
of an inefficient starting portfolio $\initp$.

Under $\boldsymbol{\pi}(t|\bp)$, the  mean and variance of $\m{y}(\mathbf{x})$ are
\begin{align}
\mathbb{E}_{\boldsymbol{\pi}(t|\bp)}\!\bigl[\mathbf{y}(\mathbf{x})\bigr]
& = \bar{\mathbf{b}}_{t,\bar{\mathbf{x}}}^{\top}\bigl(\mathbf{x}-\bar{\mathbf{x}}\bigr),\label{eq:inc-mean} \\
\Var_{\boldsymbol{\pi}(t|\bp)}\!\bigl[\mathbf{y}(\mathbf{x})\bigr]
& = \bigl(\mathbf{x}-\bar{\mathbf{x}}\bigr)^{\top}\bar{\mathbf{Q}}_{t,\bar{\mathbf{x}}}\bigl(\mathbf{x}-\bar{\mathbf{x}}\bigr).\label{eq:inc-var}
\end{align}
These two quantities are used as coefficients of the quadratic objective function introduced in Section~\ref{sec:objective}.
The tilted weights depend on the entire set of base-point P\&Ls $\pls{s}{\bp}$ through exponential reweighting, so varying $t$ changes the tilted mean and covariance.
In this sense, exponential tilting allows the resulting quadratic coefficients to retain information from the base-point distribution beyond the Gaussian case, although such features are not modeled explicitly.

\subsection{Base-point portfolio as relaxed Markowitz solution}\label{subsec:mark_and_basepoint}

The Markowitz problem consists in finding $\p{x}_{M}$ such that
\begin{equation}
\label{eq:markowitz_problem}
\p{x}_{M}=\argmin_{\mathbf{x}\in\{0,1\}^d} \left\{\objM(\pTx{\p{x}}) \right\},
\end{equation}
where
\begin{equation}
\label{eq:markowitz_obj}
\objM(\pTx{\p{x}}):=\frac{1}{2}\Var_\mathbf{m}[\pl{\pTx{\p{x}}}] - \E_\mathbf{m}[\pl{\pTx{\p{x}}}];
\end{equation}
with $\mathbf{m}$ we denote the uniform probability measure, whose entries are all equal to $1/S$. Defining
\begin{align*}
    \boldsymbol{\mu}_{\m{G}} & := \E_\mathbf{m}[\m{G}], \\
    \Sigma_{\m{G}\m{G}}  & := \mathrm{Cov}_\mathbf{m}(\m{G},\m{G}), \\
    \Sigma_{\m{G}\initp} & := \mathrm{Cov}_\mathbf{m}(\pl{\initp},\m{G}),
\end{align*}
we can derive
\begin{align*}
    \E_\mathbf{m}[\pl{\pTx{\p{x}}}] & =\E_\mathbf{m}[\pl{\initp}]+\boldsymbol{\mu}_{\m{G}}^\top\,\p{x}\,,\\
    \Var_\m{m}[\pl{\pTx{\p{x}}}] & = \Var_\m{m}[\pl{\initp}] + \p{x}^\top\Sigma_{\m{G}\m{G}}\p{x} + 2\Sigma_{\m{G}\initp}^\top\p{x}.
\end{align*}

We choose $\bar{\p{x}}$ to be the continuous solution, found through quadratic programming, of the relaxed Markowitz problem
\begin{equation}
\bar{\p{x}}=\argmin_{\mathbf{x}\in[0,1]^d} \left\{\objM(\pTx{\p{x}}) \right\}.
\end{equation}

The rationale for this choice and alternative base-point selections are discussed in Appendix \ref{app:recentering}.

\section{Quadratic Objective as a Second-Order CARA Certainty Equivalent}
\label{sec:objective}

This section introduces the tilted quadratic objective function proposed in this work: a second-order approximation of a CARA certainty equivalent.
The key idea is to optimize the metrics related to the P\&L difference relative to the chosen base-point portfolio $\bp$, while measuring both expected gain and residual dispersion under the Esscher-tilted weights.   

We proceed as follows:
(i)~we introduce the certainty equivalent as a cash-valued scalar criterion that maps P\&L distributions to comparable scores;
(ii)~we motivate the specific choice of CARA (exponential) preferences;
(iii)~we show that the exact CARA certainty equivalent is non-quadratic and derive the second-order (mean--variance) surrogate.

\subsection{Certainty equivalent as a cash-valued score}
\label{subsec:ce-criterion}

The optimization variable~$\mathbf{x}\in\{0,1\}^d$ induces a distribution of the P\&L difference across scenarios, $\m{y(\m{x})}$, under the probability measure~$\boldsymbol{\pi}(t|\bp)$ (hereafter also denoted respectively by $\m{y}$ and $\boldsymbol{\pi}$). 
To decide whether one portfolio is preferable to another, we therefore need a scalar criterion that maps each such P\&L distribution to a single cash-valued score (e.g., in euros), so that different portfolios can be compared by one number.

A standard approach to obtain such a scalar score is through expected-utility.
Rather than optimizing expected utility directly, we use its CE: the deterministic cash amount that delivers the same expected utility as the risky P\&L, providing a single monetary score that summarizes the entire distribution of P\&L difference across scenarios~\cite{gollier2001economics}.


In our context, given a utility function~$U_\gamma:\mathbb{R}\to\mathbb{R}$,
i.e., a numerical representation of the decision-maker's preferences over monetary outcomes under uncertainty, and a probability measure~$\mathbf{w}$, the certainty equivalent $\mathrm{CE}^{\mathbf{w}}_\gamma(\m{y})$ is defined as the sure cash amount~$c$ that satisfies 
\begin{equation}\label{eq:ce-def}
  U_\gamma(c) \;=\; \E_{\mathbf{w}}\!\bigl[U_\gamma(\m{y})\bigr],
\end{equation}
where $U_\gamma$ is applied to each component of the vector $\m{y}$.
Each $y_s$ is viewed as a possible outcome of a discrete random quantity on the finite scenario set $\{1,\dots,S\}$ under $\mathbf{w}$.
Maximizing $\mathrm{CE}^{\mathbf{w}}_\gamma\!\bigl(\m{y}\bigr)$ is equivalent to selecting the portfolio with the largest risk-adjusted cash equivalent value under the chosen probability measure~$\mathbf{w}$, 
i.e., the highest guaranteed cash amount that the investor would accept instead of the risky P\&L difference.
In particular, when the utility function is concave, it penalizes unfavorable outcomes more than it rewards favorable ones, even when their expected P\&L is similar; therefore, portfolios with higher downside risk have lower CE.

We use a CE-based objective function because it is
(i)~\emph{monetary}: one score in, e.g., euro;
(ii)~\emph{risk-aware}: it penalizes unfavorable outcomes through the curvature
     of~$U_\gamma$;
(iii)~\emph{compatible with scenario reweighting}: changing~$\mathbf{w}$
      (here via Esscher tilting) changes the expectation operator, without altering the optimization pipeline.



\subsection{Rationale for CARA utility}\label{subsec:why-cara}

In our setting, three operational requirements narrow the choice of the utility function.

\begin{enumerate}

\item \textbf{Cash-valued output.}
  The criterion must produce a score in the same monetary unit as the P\&L (e.g., EUR), to maintain direct comparability between different portfolios.

\item \textbf{Translation invariance.}
  Adding an amount~$c$ to the P\&L should shift the evaluation by exactly~$c$:  $\mathrm{CE}^{\mathbf{w}}_\gamma(\m{y}+c\mathbf{1}) = \mathrm{CE}^{\mathbf{w}}_\gamma(\m{y}) + c$, where $\m{1}$ is the vector whose each component is one. This ensures that the optimal solution to our optimization problem does not depend on the fixed base-point portfolio $\bp$ defined by $\m{\bar{x}}$.

\item \textbf{Well-defined for negative outcomes.}
  The P\&L difference can be negative (the new portfolio may underperform the base-point in some scenarios), so the utility must also be defined for negative values.
\end{enumerate}

\noindent
A formulation that satisfies all three requirements simultaneously is provided by the CARA family
\begin{equation}\label{eq:cara-utility}
  U_\gamma(y) \;:=\; -\exp(-\gamma\, y), \qquad y\in\RR, \qquad \gamma > 0,
\end{equation}
where the parameter~$\gamma$ is the absolute risk-aversion coefficient (units: e.g., $1/\mathrm{EUR}$).
The utility function in Eq. \eqref{eq:cara-utility} is well-defined for negative outcomes, the closed-form CE (derived in Section~\ref{subsec:ce-to-quad}) provides a cash-valued output in monetary units and the translation invariance follows from the exponential structure. 



\subsection{From exact CARA CE to the quadratic objective function}
\label{subsec:ce-to-quad}

Substituting the CARA utility~\eqref{eq:cara-utility} into the CE definition~\eqref{eq:ce-def} and solving for~$c$ yields the closed form
\begin{equation}\label{eq:cara-ce-exact}
  \mathrm{CE}^{\mathbf{w}}_\gamma(\m{y})
  \;=\;
  -\frac{1}{\gamma}\,
  \log\,\mathbb{E}_{\mathbf{w}}\!\bigl[\exp(-\gamma\, \m{y})\bigr],
\end{equation}
where the exponential function is applied component-wise.


To use the CE based on CARA utility as an objective function suitable for QUBO, we develop a quadratic approximation of Eq. \eqref{eq:cara-ce-exact}.
For a fixed tilted measure $\boldsymbol{\pi}(t|\bp)$, we consider $\mathrm{CE}^{\boldsymbol{\pi}(t|\bp)}_\gamma(\m{y})$ as a function of the risk-aversion parameter $\gamma$ and use the usual local expansion for small risk aversion around $\gamma=0$, i.e., around the risk-neutral limit in which
the certainty equivalent reduces to the expected P\&L; see, e.g.,
\cite{gollier2001economics,gerber1998utility}.
Indeed,
\[
\mathbb{E}_{\boldsymbol{\pi}}\!\left[\exp(-\gamma \m{y})\right]
=
1-\gamma\,\mathbb{E}_{\boldsymbol{\pi}}[\m{y}]
+\frac{\gamma^{2}}{2}\,\mathbb{E}_{\boldsymbol{\pi}}[\m{y}^{2}]
+\mathcal{O}(\gamma^{3})
\]
where $\m{y}^2$ is the vector whose components are the square of the components of $\m{y}$.
Applying $\log(1+v)=v-\frac{v^{2}}{2}+\mathcal{O}(v^{3})$ with
\[
v=
-\gamma\,\mathbb{E}_{\boldsymbol{\pi}}[\m{y}]
+\frac{\gamma^{2}}{2}\,\mathbb{E}_{\boldsymbol{\pi}}[\m{y}^{2}]
+\mathcal{O}(\gamma^{3}),
\]
gives
$v^{2}
=
\gamma^{2}\bigl(\mathbb{E}_{\boldsymbol{\pi}}[\m{y}]\bigr)^{2}
+\mathcal{O}(\gamma^{3})$,
and therefore
\begin{multline}\label{eq:log_E_exp_II_ord}
\log \mathbb{E}_{\boldsymbol{\pi}}\!\left[\exp(-\gamma \m{y})\right] = \\
-\gamma\,\mathbb{E}_{\boldsymbol{\pi}}[\m{y}] +\frac{\gamma^{2}}{2}\,\mathbb{E}_{\boldsymbol{\pi}}[\m{y}^{2}] 
 -\frac{\gamma^{2}}{2}\bigl(\mathbb{E}_{\boldsymbol{\pi}}[\m{y}]\bigr)^{2} +\mathcal{O}(\gamma^{3}) =\\
 -\gamma\,\mathbb{E}_{\boldsymbol{\pi}}[\m{y}] +\frac{\gamma^{2}}{2}\Var_{\boldsymbol{\pi}}(\m{y}) +\mathcal{O}(\gamma^{3}).
\end{multline}
Substituting \eqref{eq:log_E_exp_II_ord} into \eqref{eq:cara-ce-exact} and omitting $\mathcal{O}(\gamma^2)$, we obtain
\begin{equation}\label{eq:ce-mv}
  \mathrm{CE}^{\boldsymbol{\pi}(t|\bp)}_{\gamma}
    \!\bigl(\m{y}(\mathbf{x})\bigr)
  \;\approx\;
  \mathbb{E}_{\boldsymbol{\pi}(t|\bp)}\!\bigl[\m{y}(\mathbf{x})\bigr]
  \!-\!
  \frac{\gamma}{2}\,
  \Var_{\boldsymbol{\pi}(t|\bp)}
    \!\bigl[\m{y}(\mathbf{x})\bigr].
\end{equation}
By maximizing this quantity, configurations with a higher tilted mean and lower tilted variance are rewarded, in coherence with the Markowitz approach.
Higher-order distributional features (skewness, kurtosis, \ldots)
are not explicitly modeled in the quadratic surrogate;
their influence enters only indirectly through the tilted
weights~$\boldsymbol{\pi}(t|\bp)$ used to compute the two moments in Eq. \eqref{eq:ce-mv}.

The expansion is locally accurate when~$\gamma$ is small relative to the scale of~$\m{y}$ components, as in the case of our experimental settings.
Accordingly, we use the resulting tilted mean--variance form as a
quadratic surrogate of the exact CARA certainty
equivalent under the tilted measure. The resulting objective is a standard and self-consistent quadratic criterion for any $\gamma>0$: the second-order expansion motivates the functional form and clarifies its economic content (reward versus penalty structure), but the quadratic surrogate does not rely on $\gamma$ being small in order to define a well-posed optimization problem.

It is important to note that the framework involves two parameters with distinct roles:
\begin{itemize}
\item
  the tilt~$t$ defines the measure
  $\boldsymbol{\pi}(t|\bp)$, i.e., which scenarios are emphasized
  (upside for $t>0$, downside for $t<0$);
\item
  the risk-aversion coefficient~$\gamma$ controls the tilted
 mean--variance trade-off once the measure is fixed:
  larger~$\gamma$ emphasizes the second term of Eq. \eqref{eq:ce-mv}.
\end{itemize}

In our experiments, we couple the two parameters via $\gamma = |t|$. Substituting
the tilted mean of Eq.~\eqref{eq:inc-mean} and variance of Eq.~\eqref{eq:inc-var}
into Eq.~\eqref{eq:ce-mv} yields the objective function to be minimized
\begin{equation}\label{eq:centered-obj}
\begin{aligned}
  \obj_t(\pTx{\p{x}}) := {} & -\,\bar{\mathbf{b}}_{t,\bp}^{\top}
    \bigl(\mathbf{x}-\bar{\mathbf{x}}\bigr) \\
  & + \frac{|t|}{2}\,
  \bigl(\mathbf{x}-\bar{\mathbf{x}}\bigr)^{\top}
  \bar{\mathbf{Q}}_{t,\bp}\,
  \bigl(\mathbf{x}-\bar{\mathbf{x}}\bigr),
\end{aligned}
\end{equation}
where the tilted mean vector
$\bar{\mathbf{b}}_{t,\bp}$ and the covariance matrix
$\bar{\mathbf{Q}}_{t,\bp}$ are the weighted sample
averages defined in Section~\ref{subsec:stressed-moments}.

Under this coupling, more severe tilting (larger~$|t|$) is accompanied
by a stronger dispersion emphasis,
reflecting the view that a practitioner choosing a more extreme scenario reweighting may pay more attention to risk. 
Moreover, the coupling $\gamma=|t|$ does not imply symmetry between positive and negative tilts, as 
the measure $\boldsymbol{\pi}(t|\bp)$ depends on the sign of $t$ through the exponential reweighting of
scenarios. 
Consequently, the tilted moments $\bar{\mathbf{b}}_{t,\bp}$ and $\bar{\mathbf{Q}}_{t,\bp}$ generally differ for $t$ and $-t$, and the resulting optimization problems need not produce symmetric solutions.

In our experiments, the tilt grid is restricted to moderate values of $|t|$, to remain in the same qualitative regime as the second-order surrogate in Eq.~\eqref{eq:ce-mv}.




\section{QUBO formulation}
\label{sec: qubo}

In Section \ref{subsec:solver-coeffs} we present a QUBO formulation of the tilted quadratic objective function defined in Eq.~\eqref{eq:centered-obj}, and in the subsequent Sections (\ref{subsec:Instrument-Type Constraint}, \ref{subsec:Unique Notional Value per Instrument}, \ref{subsec:Greek Constraints}) we then show how to incorporate as penalty terms the business constraints defined in Section \ref{sec:rw-riskaware}.

\subsection{QUBO objective function}
\label{subsec:solver-coeffs}

We begin by deriving a QUBO representation of the tilted quadratic objective function defined in Eq.~\eqref{eq:centered-obj}, which serves as the foundation for the optimization framework.

A QUBO problem consists of finding $\mathbf{x}_{\mathrm{opt}}$ such that
\begin{equation}\label{eq:QUBO-obj}
\mathbf{x}_{\mathrm{opt}} = \argmin_{\mathbf{x}\in\{0,1\}^d} \left\{ \mathbf{x}^{\top} \mathbf{Q}_{\mathrm{obj}} \mathbf{x} \right\}.
\end{equation}
where $\mathbf{Q}_{\mathrm{obj}} \in \RR^{d\times d}$ is a symmetric matrix. 

In our context, $\mathbf{x} \in \{0,1\}^d$ is a binary decision vector whose component $x_u$ takes the value $1$ if a specific UEI and its notional value is included in the EOS and $0$ otherwise (as discussed in Section \ref{subsec:pl_decomposition}). Up to constant terms, the optimization problem based on the objective function $\obj_t(\pTx{\p{x}})$ in Eq.~\eqref{eq:centered-obj} can be written, by expanding and collecting linear and quadratic terms, as
\begin{equation}
\label{eq:quadratic_problem}
\p{x}_{t} = \argmin_{\mathbf{x}\in\{0,1\}^d} \left\{\mathbf{x}^{\top}\left(\frac{|t|}{2}\,\bar{\mathbf{Q}}_{t,\bp} - \operatorname{diag}(\mathbf{b})\right)\mathbf{x}\right\}\,,
\end{equation}
that satisfies the formulation of Eq. \eqref{eq:QUBO-obj}, where
\begin{equation}\label{eq:QP-coeff}
    \mathbf{b} =\bar{\mathbf{b}}_{t,\bp}
     + |t|\,\bar{\mathbf{Q}}_{t,\bp}\,\bar{\mathbf{x}}\,.
\end{equation} 
In this formulation, minimizing the objective function can be interpreted as selecting specific UEIs, together with a notional value, to be added to the initial portfolio $\initp$ to obtain an optimized final portfolio configuration $\pT$.


\subsection{EOS composition constraint}
\label{subsec:Instrument-Type Constraint}
Let $I$ denote the set of indices corresponding to instruments of a given type $i$ and associated with the same underlying asset. If $n_i$ is the maximum allowed number of instruments of type $i$, the constraint can be written as
\[
\sum_{u \in I} x_u = n_i\,.
\]
It is worth noting that this equality implicitly imposes an inequality constraint. In fact, the $x_u$ in the above expression may correspond to a notional value equal to zero, which is equivalent to not selecting an instrument of the given type.

This constraint can be mapped into the QUBO framework through the quadratic penalty term
\begin{equation}\label{constr:instrument}
\left( \sum_{u \in I} x_{u} - n_i \right)^2,
\end{equation}
which is added to the objective function for each underlying asset and type of instrument $i$, scaled by an appropriate penalty coefficient.

\subsection{UEI quantities constraints}
\label{subsec:Unique Notional Value per Instrument}
Since the choice of a notional value for a UEI corresponds to set to $1$ only a single binary variable among those in $\mathbf{x}$ associated with that UEI, we must ensure that
\[
\sum_{u \in U} x_{u} \le 1 \quad \forall \: U,
\]
where $U$ denotes the set of indices corresponding to different notional values of the same UEI. The total number of such sets $U$ is $\nU$. The inequality is fundamental because enforcing equality would encourage the addition of the UEI into the EOS, thereby creating an inconsistency with the EOS composition constraint. With this formulation, the discretization of the notional values is naturally enforced.

This constraint can be formulated as a QUBO penalty term as follows
\begin{equation}\label{constr:notional_value}
\sum_{u=1}^{\nN-1}
x_{u}
\sum_{u'=u+1}^{\nN}
x_{u'}
\quad \forall \: U\,.
\end{equation}

\subsection{Greeks constraints}
\label{subsec:Greek Constraints}
The Greeks measure the sensitivity of the portfolio value to small changes in underlying risk factors. The Greeks of our interest are Delta ($\Delta$), Vega ($\Vega$), and Gamma ($\Gamma$).
In practice, they encode the risk profile of the initial portfolio $\initp$ decided by traders, according to their market view.
The Greeks constraints prescribe their maximum allowed variation, to control the change in the risk profile of the optimized portfolio $\mathcal{T}$ with respect to 
$\initp$.


Let $\mathcal{G}_\initp$ and $\mathcal{G}_\pT$ denote the value of one of the Greeks of the portfolios $\initp$ and $\pT$, respectively, $\text{tol}_{\mathcal{G}}$ the maximum allowed percentage variation of $\mathcal{G}_\initp$, and $g_u$ the Greek value associated with the $u$-th pair UEI-notional value. The constraint is
\begin{equation}
|\mathcal{G}_\pT - \mathcal{G}_\initp| = \left| \sum_{u} g_ux_u \right| < C_{\mathcal{G}}
\label{eq:greek}
\end{equation}
where $C_\mathcal{G} :=\text{tol}_{\mathcal{G}} \, |\mathcal{G}_\initp|$.

We consider two strategies to embed this constraint in the formulation: slack variables and Lagrangian relaxation.
The first approach adopts the \textbf{slack variable} technique \cite{qubo} to reformulate the inequality constraint as a quadratic penalty term
\begin{equation}
\label{eq:slack_penalty}
\left( \sum_{u} g_u x_u - C_\mathcal{G} + V \right)^2,
\end{equation}
where $V = \sum_{j=0}^{m}2^j v_j$ is a slack variable, discretized in binary form, whose resolution controls the desired precision of the Greek constraint. The binary slack variable components $v_j$ must be added to the decision variables of the QUBO problem, extending the solution space by $m=\lceil\log_2(2C_\mathcal{G}+1)\rceil$.
Besides the increased dimensionality of the QUBO problems, note that we have three Greek constraints for each problem instance, and therefore three sets of binary slack variables. Expressing the constraints in this way introduces additional penalty parameters whose appropriate values are not trivial to determine to reach optimality.

For this reason, we introduce an alternative method to incorporate these constraints into the QUBO objective function, based on \textbf{Lagrangian relaxation} \cite{fisher1981lagrangian}. 
This approach relies on treating the constraint as a \emph{complicating constraint} and incorporating it into the objective function via a multiplier.
To do so, we need to express the constraint as $l(\p{x})\leq 0$ for some function $l$ that must be quadratic to be embedded in the objective function. The absolute value in Eq. \eqref{eq:greek} is addressed by squaring the constraint, yielding
$$
l(\p{x}) := \left(\frac{\sum_u g_u x_u}{C_\mathcal{G}}\right)^2 - 1,
$$
where we divide by $C_\mathcal{G}$ for numerical stability.
Now, we define the final objective function as
\begin{equation}\label{eq:obj_func_lambda}
    \obj_{\lambda}(\pTx{\m{x}})=\p{x}^{T} \m{Q}_{\mathrm{pen}} \p{x} + \lambda \, l(\p{x})
\end{equation}
for some positive multiplier $\lambda$ that maximizes
\begin{equation}\label{eq:sub_dual_problem}
f(\lambda) := \min_{\p{x} \in \{0,1\}^d}
\obj_{\lambda}(\pTx{\m{x}})\,.
\end{equation}
Here, the matrix $\mathbf{Q}_{\mathrm{pen}}$ incorporates both the objective component $\mathbf{Q}_{\mathrm{obj}}$ and the penalty terms associated with the constraints discussed in Sections \ref{subsec:Instrument-Type Constraint} and \ref{subsec:Unique Notional Value per Instrument}.
To find an adequate value for $\lambda$, we adopt an iterative gradient-based method \cite{fisher1981lagrangian}: at the $i$-th iteration, given some $\lambda=\lambda^{(i)}$, we use a QUBO solver to find $\p{x}^{(i)}\in\{0,1\}^d$ that minimizes $\obj_{\lambda^{(i)}}(\pTx{\m{x}})$.
The function $l$ evaluated in $\p{x}^{(i)}$ represents a valid supergradient of $f$ in $\lambda^{(i)}$, since it can be shown that
$$
f(\lambda) - f(\lambda^{(i)}) \leq l(\p{x}^{(i)})(\lambda - \lambda^{(i)})\,,
$$
for every $\lambda>0$. So, the next multiplier is 
$$\lambda^{(i+1)} = \max{(0, \lambda^{(i)} + \mu^{(i)} l(\p{x}^{(i)})) }$$
where $\mu^{(i)}$ is a properly chosen iteratively decreasing factor (see Section \ref{sec:experiments}).

The same approach can be extended to manage all Greeks, by introducing a Lagrangian multiplier ($\lambda_\Delta$, $\lambda_\Gamma$ and $\lambda_\Vega$) and a constraint function ($l_\Delta$, $l_\Gamma$ and $l_\Vega$) for each Greek. 
Hence, the complete Lagrangian dual problem is
$$
    \max_{\lambda_\Delta,\lambda_\Gamma,\lambda_\Vega \ge 0} \; \min_{\p{x} \in \{0,1\}^d}
\left(
\p{x}^{T} \m{Q}_{\mathrm{pen}} \p{x} + \sum_{k\in\{\Delta,\Gamma,\Vega\}} \lambda_k \, l_k(\p{x})
\right).
$$
Note that Lagrangian multipliers are optimized simultaneously.


\section{Experiments}\label{sec:experiments}
In our experimental analysis, we first focused on validating the proposed tilting framework, assessing the impact that different tilt values have on the optimization process. We adopted the constrained formulation of the optimization problem and a reduced dataset to provide greater guaranties of obtaining a feasible, optimal solution using MIQP.

The results were analyzed to identify a subset of representative tilts used to test the QUBO formulation. Each constrained tilted model was translated into its equivalent unconstrained representation and executed using MIQP and the Simulated Bifurcation quantum-inspired method.

To assess scalability, we further investigated the computational tractability of larger problem instances under MIQP. Using the full dataset and the Markowitz formulation, we evaluated the execution times of MIQP solvers (for both constrained and unconstrained versions of the problem) and of the Simulated Bifurcation heuristic.

\subsection{Dataset and problem setting}\label{subsec:datasets}
For our tests, we adopted the same dataset and settings as used in \cite{bianchetti2025riskawaretradingportfoliooptimization}. The dataset consists of 8 stocks and 5 stocks indexes. For each underlying, the unique eligible instruments are identified by a choice of instrument type (call, put, future, or stock), strike, and maturity. See \cite[Table 3]{bianchetti2025riskawaretradingportfoliooptimization} for details. To replicate the setting, we build $\pT$ by selecting at most two options and one stock/future per underlying asset. For sensitivities, we restrict our experiments to a maximum fixed percentage variation $\text{tol}_{\Delta} = \text{tol}_{\Gamma} = \text{tol}_{\Vega}= 100\%$.

The number $\nN$ of discretized notional values for each underlying is $21$, corresponding to $10$ positive amounts, $10$ negative and $0$, all equally spaced within the same interval centered at $0$. In other words, notional values are discretized using a step that is a tenth of the maximal allowed notional value.
Most of our experiments were performed replicating the small-sized application setting of \cite{bianchetti2025riskawaretradingportfoliooptimization} that limits the dataset to only one underlying:  the Euro Stoxx 50 Index (.STOXX50E). Hereafter, the subset of all unique eligible instruments associated to this underlying is named \emph{reduced} dataset. The combinations of instrument type, strike, and maturity for .STOXX50E are $42$, yielding a dimension $d=42\times 21=882$.
However, some final experiment (see Section \ref{subsec:scaling_up}) has been performed using the complete dataset, hereafter named \emph{full} dataset, for a total of $556$ UEIs and a global dimension $d=21\times 556=11\,676$.

Actually, for numerical reasons, the dataset was pre-processed by dividing all entries in $\m{G}$ by the standard deviation $\sigma_\m{G}$, which is $19\,051.49$ for the reduced dataset and $18\,473.17$ for the full dataset.
Adopting the proposed framework using a normalized dataset is formally equivalent to using the original data and scaling tilts by the same factor $\sigma_\m{G}$. The reader should be aware that the tilt values reported in this paper refer to the normalized dataset.

Finally, in accordance with the Markowitz formulation (see Section \ref{subsec:mark_and_basepoint}), the base probability measure $\m{w}$ applied to scenarios before tilting is the uniform distribution $\m{m}$.

\subsection{Tilting framework validation}\label{subsec:validation}
\begin{figure}
    \centering
    \includegraphics[width=1\linewidth, trim=0.3cm 0cm 0cm 0cm, clip]{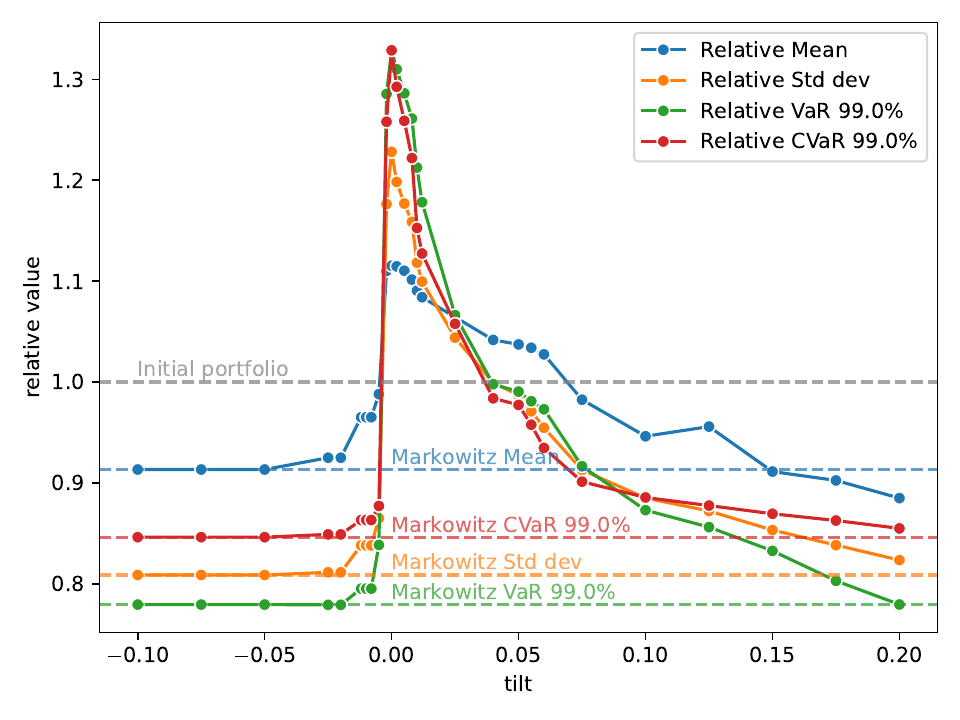}
    \caption{Plot of relative (with respect to the initial portfolio) mean, standard deviation, VaR and CVaR (at $99\%$ level), in dependence of tilts values. For the sake of clarity, VaR and CVaR at $97.5\%$ level are not presented, since they behave very similarly to VaR and CVaR at $99\%$.}
    \label{fig:mean_std}
\end{figure}

\definecolor{mark_color}{rgb}{0, 0, 1}
\definecolor{init_color}{HTML}{FFA500}
\definecolor{vs_init_color}{HTML}{008000}
\definecolor{SR_color}{rgb}{1, 0, 0}

We considered all tilts on a grid with a step size of $0.025$, ranging from $-0.1$ to $0.2$, plus some more tilt in some sub-interval that required further investigation.
For each tilt in this set, we employed \Gurobi{} to solve the problem defined in Eq. \eqref{eq:quadratic_problem}, 
subject to the constraints described in Section \ref{sec: qubo} handled explicitly and not as penalties.
As already mentioned in Section~\ref{subsec:mark_and_basepoint}, the base-point $\bar{\mathbf{x}}$ required by the model is chosen to be the continuous solution of the mean--variance problem. In order to find such a solution, we adopted CVXOPT through \textsc{qpsolvers} \cite{qpsolvers}, which has proven to be very fast and essentially negligible in terms of time during the execution of the entire process. 

Each solution $\p{x}_t$ (one for each tilt $t$) was used to build a new portfolio $\pT=\pTx{\p{x}_t}$. The P\&L distribution of each portfolio was then primarily evaluated in terms of its mean, standard deviation, VaR, and CVaR at $99\%$ and $97.5\%$ levels.
Figure \ref{fig:mean_std} shows the trends of the mean (in blue), standard deviation (in orange), VaR (in green) and CVaR at $99\%$ (in red) of the solutions as the tilt varies.
The metrics values are normalized by dividing by the corresponding value of the initial portfolio $\initp$.
Therefore, a normalized mean greater than one indicates an improved performance relative to $\initp$. For VaR and CVaR, the interpretation requires care. In our convention, these risk metrics are negative, and an improvement corresponds to values closer to zero (smaller absolute magnitude).
Hence, if both the VaRs of $\initp$ and $\pT$ are $\text{VaR}_{\initp}<0$ and  $\text{VaR}_\pT<0$, the normalized ratio $\text{VaR}_\pT/\text{VaR}_{\initp}$ is positive and an improved (less negative) $\text{VaR}_\pT$ generates a ratio below one (and analogously for CVaR).
In this figure, the initial portfolio metrics are highlighted by the gray dashed line, and all have values equal to $1$. 
Instead, colored dashed lines correspond to the same metrics evaluated on the solution of the Markowitz problem (see Eq. \eqref{eq:markowitz_problem}).

All metrics shown exhibit similar behavior, flattening toward the edges of the grid and displaying a cusp-like maximum at zero.
This latter feature was expected \textit{a priori} and is due to the factor $|t|$ that multiplies the quadratic component: when $t$ is zero, only the linear part, i.e. the mean, is optimized, as the quadratic risk term does not contribute.

Starting from zero and moving toward negative tilts, we observe a steep slope. This is because the quadratic component gradually gains importance, and negative tilts emphasize adverse scenarios; consequently, the optimization process focuses on minimizing losses in these scenarios rather than maximizing returns.
Conversely, as the weight of the quadratic component continues to increase for positive tilts, these emphasize favorable scenarios where achieving higher returns is more advantageous than simply reducing risk. Therefore, the slope is less steep on this side.

The flattening observed at the edges is expected, since extreme tilts effectively select a very limited number of scenarios through exponential tilting. This becomes particularly intuitive when considering the limiting behavior: for very large tilts, the weighting scheme concentrates on a single scenario, and the solution to the corresponding optimization problems therefore becomes invariant.
For negative tilts, flattening occurs rapidly and the solution converges to the Markowitz optimum. This behavior was unexpected and is likely related to the specific characteristics of the dataset used in the experiments. In practice, solving the Markowitz problem on this dataset appears equivalent to solving it on an effectively single-scenario dataset, namely, the one emphasized by exponential tilting.

Moreover, it emerged experimentally that with this particular reduced dataset, the quadratic part in the Markowitz objective function is much more significant than the linear part and, since we are constrained to choose just a limited number of UEIs to replicate the setting of \cite{bianchetti2025riskawaretradingportfoliooptimization}, solving the Markowitz problem is indeed equivalent to minimize the variance.
This explains why, varying the tilts, we cannot obtain any solution reducing the standard deviation with respect to the Markowitz solution.

%

\begin{figure*}[t]
    \centering
    \includegraphics[width=1\linewidth]{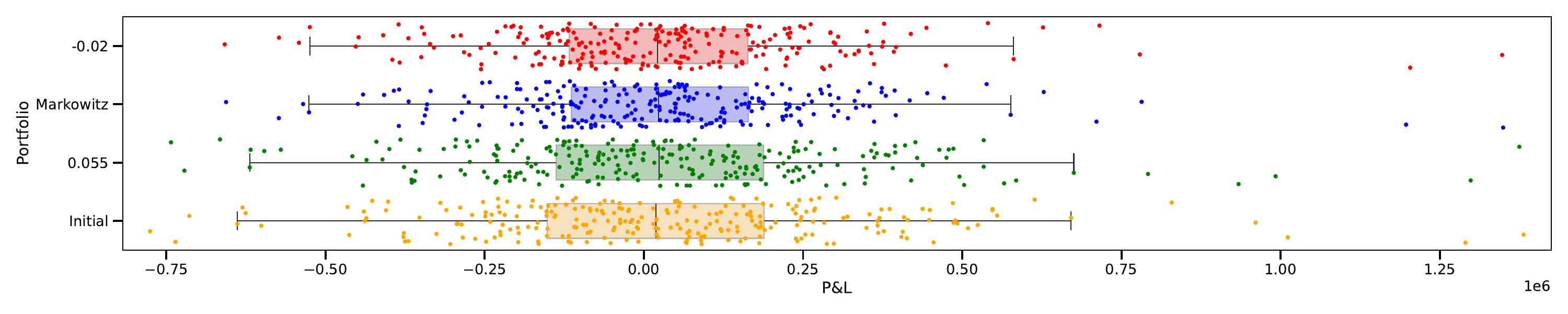}
    \caption{Box plots comparing the P\&L distributions of the initial portfolio (orange), the Markowitz solution (blue), and the solutions for tilts $0.055$ (green) and $-0.02$ (red), illustrating how different tilts capture different distribution features.}
    \label{fig:boxplot}
\end{figure*}

\begin{figure*}[t]
    \centering
    \includegraphics[width=1\linewidth]{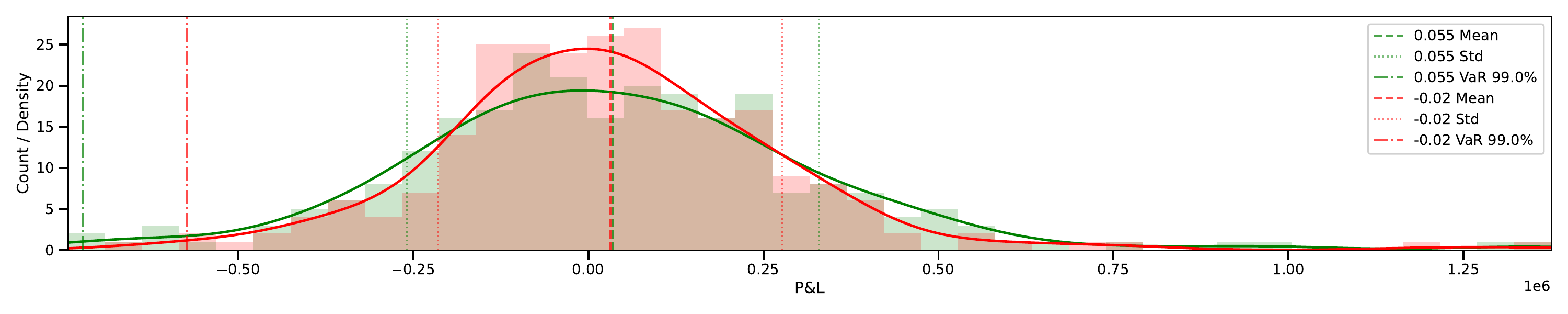}
    \caption{Histograms comparing the P\&L distribution of the solution for tilts $0.055$ (in green) and $-0.02$ (in red). Vertical lines report the mean, the VaR at $99.0\%$ and the mean $\pm$ the standard deviation. It should be noted that while the former shows a higher mean (i.\ e.\ expected return), the latter reduces risk-related metrics.}
    \label{fig:distribution_0_055}
\end{figure*}

In summary, the figure shows that varying the tilts produces solutions targeting different P\&L objectives.
This can also be observed in Figures \ref{fig:boxplot} and \ref{fig:distribution_0_055}, where solutions with tilts $-0.02$ and $0.055$ exhibit markedly different, but interesting, features.
Indeed, we identified two neighborhoods around these values, namely $[0.04,0.06]$ and $[-0.025,-0.005]$, that preserve this behavior and are therefore worth further investigation.

Let us start from the interval $[0.04, 0.06]$. The solutions for these tilts differ from the Markowitz one, and in particular they perform worse with respect to risk-related metrics (such as standard deviation, VaR and CVaR), whereas the Markowitz solution delivers a smaller return.
However, they perform better than the initial portfolio with respect to every considered metric.
This fact can be observed in Table \ref{tab:details} and Figure \ref{fig:mean_std}, where the blue points of the interval exceed $1$, while all other points lie below $1$.

The solutions found for tilts in the interval $[-0.025, -0.005]$, on the other hand, are intermediate between the initial portfolio, which is more return-oriented, and the mean--variance solution, which is more conservative. However, they are characterized by a higher mean-to-standard deviation ratio. Indeed, these solutions also exhibit good mean-to-VaR ratios, as shown in Table \ref{tab:details}. The table also reports the performances obtained through the RATPO algorithm, which slightly outperforms others with respect to the mean-to-VaR ratio at the confidence level $99\%$, this is expected since both metrics contribute to the construction of the RATPO fractional objective function. However, it performs worse according to all other metrics than the solutions in the interval.

\definecolor{mark_color}{rgb}{0, 0, 1}
\definecolor{init_color}{HTML}{FFA500}
\definecolor{vs_init_color}{HTML}{008000}
\definecolor{SR_color}{rgb}{1, 0, 0}

\subsection{Validating QUBO formulation}

\begin{table*}[t]
    \centering
        \begin{tabular}{lrrrrrrrrr}
        \toprule
        Instance & Mean & Std dev & VaR 99.0\% & CVaR 99.0\% & VaR 97.5\% & CVaR 97.5\% & Mean/Std & Mean/VaR99\% & Mean/VaR97.5\% \\
        \midrule
            \rowcolor{init_color!30}
            Initial & 33\,918 & 302\,757 & -735\,750 & -775\,588 & -714\,031 & -755\,669 & 0.1120 & -0.0461 & -0.0475 \\
            RATPO & 32\,634 & 255\,585 & -582\,060 & -677\,105 & -581\,533 & -629\,582 & 0.1277 & -0.0561 & -0.0561 \\
            \rowcolor{mark_color!30}
            Markowitz & 30\,974 & 244\,804 & -573\,327 & -656\,230 & -535\,283 & -614\,779 & 0.1265 & -0.0540 & -0.0579 \\
            t=-0.1 & 30\,974 & 244\,804 & -573\,327 & -656\,230 & -535\,283 & -614\,779 & 0.1265 & -0.0540 & -0.0579 \\
            t=-0.075 & 30\,974 & 244\,804 & -573\,327 & -656\,230 & -535\,283 & -614\,779 & 0.1265 & -0.0540 & -0.0579 \\
            t=-0.05 & 30\,974 & 244\,804 & -573\,327 & -656\,230 & -535\,283 & -614\,779 & 0.1265 & -0.0540 & -0.0579 \\
            \textbf{t=-0.025} & {31\,371} & {245\,604} & {-573\,145} & {-658\,376} & {-541\,731} & {-615\,761} & \textbf{0.1277} & \textbf{-0.0547} & \textbf{-0.0579} \\
            \rowcolor{SR_color!30} 
            \textbf{t=-0.02} & {31\,371} & {245\,604} & {-573\,145} & {-658\,376} & {-541\,731} & {-615\,761} & \textbf{0.1277} & \textbf{-0.0547} & \textbf{-0.0579} \\
            \textbf{t=-0.012} & 32\,734 & 253\,705 & -584\,977 & -669\,412 & -575\,860 & -627\,194 & \textbf{0.1290} & \textbf{-0.0560} & \textbf{-0.0568} \\
            \textbf{t=-0.01} & 32\,734 & 253\,705 & -584\,977 & -669\,412 & -575\,860 & -627\,194 & \textbf{0.1290} & \textbf{-0.0560} & \textbf{-0.0568} \\
            t=0.0 & 37\,825 & 371\,832 & -977\,771 & -1\,030\,644 & -808\,508 & -1\,004\,207 & 0.1017 & -0.0387 & -0.0468 \\
            t=0.025 & 36\,100 & 316\,066 & -784\,533 & -820\,297 & -744\,161 & -802\,415 & 0.1142 & -0.0460 & -0.0485 \\
            \textbf{t=0.04} & \textbf{35\,333} & \textbf{302\,494} & \textbf{-734\,122} & \textbf{-763\,045} & \textbf{-713\,917} & \textbf{-748\,584} & 0.1168 & -0.0481 & -0.0495 \\
            \textbf{t=0.05} & \textbf{35\,182} & \textbf{298\,851} & \textbf{-728\,705} & \textbf{-757\,940} & \textbf{-696\,159} & \textbf{-743\,323} & {0.1177} & {-0.0483} & {-0.0505} \\
            \rowcolor{vs_init_color!30} 
            \textbf{t=0.055} & \textbf{35\,068} & \textbf{293\,951} & \textbf{-721\,733} & \textbf{-742\,723} & \textbf{-665\,803} & \textbf{-732\,228} & {0.1193} & {-0.0486} & {-0.0527} \\
            \textbf{t=0.06} & \textbf{34\,845} & \textbf{288\,977} & \textbf{-715\,806} & \textbf{-724\,739} & \textbf{-638\,419} & \textbf{-720\,273} & 0.1206 & -0.0487 & -0.0546 \\
            t=0.075 & 33\,322 & 276\,354 & -674\,291 & -698\,916 & -597\,866 & -686\,603 & 0.1206 & -0.0494 & -0.0557 \\
            t=0.1 & 32\,137 & 268\,027 & -643\,121 & -686\,775 & -581\,849 & -664\,948 & 0.1199 & -0.0500 & -0.0552 \\
            t=0.125 & 32\,417 & 264\,024 & -629\,867 & -680\,532 & -578\,225 & -655\,199 & 0.1228 & -0.0515 & -0.0561 \\
            t=0.15 & 30\,905 & 258\,297 & -612\,509 & -674\,232 & -576\,940 & -643\,370 & 0.1196 & -0.0505 & -0.0536 \\
            t=0.175 & 30\,607 & 253\,780 & -590\,695 & -669\,025 & -576\,633 & -629\,860 & 0.1206 & -0.0518 & -0.0531 \\
            t=0.2 & 30\,012 & 249\,288 & -573\,406 & -662\,992 & -572\,541 & -618\,199 & 0.1204 & -0.0523 & -0.0524 \\
        \end{tabular}
    \caption{Metrics of a selection of solutions found by the proposed framework compared to mean--variance solution, RATPO solution and the initial portfolio. Here, all columns, but the last three that are dimensionless, are in Euros and represent actual P\&L figures. Solutions for tilts in the interval $[0.04,0.06]$ outperform the initial portfolio with respect to all displayed metrics. Interval $[-0.025, -0.005]$ instead presents solutions having high mean-standard deviation and mean-VaR ratios.
    Highlighted solutions are represented also on Figures \ref{fig:boxplot} and \ref{fig:distribution_0_055}.}
    \label{tab:details}
\end{table*}

\begin{table*}[t]
\centering
\begin{tabular}{l cc cc cc}
\toprule
 & \multicolumn{2}{c}{\Gurobi\_Q} & \multicolumn{2}{c}{\Gurobi{} MIQP} & \multicolumn{2}{c}{Simulated Bifurcation} \\
\cmidrule(lr){2-3} \cmidrule(lr){4-5} \cmidrule(lr){6-7}
Instance & time & optimality gap & time & optimum & time & optimality gap \\
\midrule
Markowitz (reduced) & 9.6 min & 0.00\% & $<10$s & -43.5591 & $>$1.5 h & unfeasible solution \\
$t=-0.02$  & 20 s & 1.77\%  & $<10$s & -0.7100  & 30 min & 0.00\% \\
$t=-0.005$ & 11.3 min & 0.00\% & $<10$s & -0.1368 & $>$1.5 h & unfeasible solution  \\
$t=0.04$   & 4.5 min & 0.00\% & $<10$s & -0.0710 & 17 min & 0.00\% \\
$t=0.055$  & 6.6 min & 0.00\% & $<10$s & -0.0847 & 36 min & 1.50\% \\
Markowitz (full) & 2.8 min & 0.85\% &  15 min & -117.47* & $>$1.5 h & unfeasible solution \\
\bottomrule
\end{tabular}
\caption{Comparison of execution times of the constrained optimization problem (\Gurobi{} MIQP) and the unconstrained versions using Lagrangian relaxation  (\Gurobi\_Q and Simulated Bifurcation).
The optimum column displays the optimal value if available, otherwise the best known bound*, found with the \Gurobi{} MIQP solver.
The optimality gap, for \Gurobi\_Q and Simulated Bifurcation, is defined as the relative difference between the best objective value found and the optimum of \Gurobi{} MIQP.
Times reported for \Gurobi\_Q/Simulated Bifurcation, correspond to the execution time required to obtain the solution used in the optimality gap evaluation; however, the solvers were allowed to run for longer ($\approx30/100$ min).}
\label{tab:running_times}
\end{table*}

Once we have validated the theoretical tilting framework, we solved again the problem in the QUBO formulation for some of the most interesting  tilts: $-0.005$, $-0.02$, $0.04$ and $0.055$. 

We compared the two strategies for handling the Greeks constraints presented in Section \ref{subsec:Greek Constraints}:
a slack variable encoding and a Lagrangian relaxation approach, in the latter case we used an update step $\mu^{(i)} = 1/\sqrt{i}$ satisfying the standard convergence requirements of the supergradient method \cite{fisher1981lagrangian}.

We tested both methods in combination with two different QUBO solvers: \Gurobi{}'s wrapper for QUBO formulations \cite{gurobiq}, hereafter named \Gurobi\_Q, and Simulated Bifurcation \cite{Ageron_Simulated_Bifurcation_SB_2023}. Furthermore, as a baseline for optimality, we relied on the constrained formulation of the MIQP problem, discussed in Section \ref{subsec:validation}.

The experiments were conducted on two Ubuntu 22.04.3 LTS machines equipped with AMD EPYC 7\,763 CPUs. 
The \Gurobi{} solver was run on a CPU-only machine (Machine~A) with 15 CPU cores and 57~GB RAM, 
while the Simulated Bifurcation solver was executed on a GPU-enabled machine (Machine~B) with 
7 CPU cores, 28~GB RAM, and an NVIDIA A30 GPU.

The slack variable formulation proved to be ineffective in the tested instances. Despite exploring multiple values for the penalty coefficients associated with the Greeks constraints and attempting to scale them with respect to the objective magnitude, the QUBO solvers were unable to obtain satisfactory solutions within a time limit of $1$ hour. In particular, \Gurobi\_Q\ consistently returned suboptimal solutions, while Simulated Bifurcation failed to find feasible solutions. A detailed report of the results obtained with this methodology can be found in Appendix \ref{app:slack}.

On the other hand, Lagrangian relaxation proved to be quite efficient, obtaining the optimal solutions for all tilts except $t=-0.02$, 
when solving with \Gurobi\_Q. The solutions were found in a few minutes, the longest run being for $0.055$ which took less than 15 minutes (see Table \ref{tab:running_times}). The Simulated Bifurcation solver proved to be less reliable and exhibited convergence behavior that was difficult to control. In general, it underperformed \Gurobi\_Q in most cases, with the exception of $t=-0.02$, for which it identified an optimal solution that \Gurobi \_Q \ did not obtain. In our experiments, we considered up to $200$ Lagrangian dual iterations; each QUBO subproblem was solved with a time limit of $10$ seconds using \Gurobi\_Q and $30$ seconds using the Simulated Bifurcation solver.

Overall, the best QUBO configuration was Lagrangian relaxation combined with \Gurobi\_Q,
which recovered all optimal solutions except for one tilted instance in the reduced dataset. See Appendix \ref{app:lagr} for further details.

\subsection{Scaling-up}\label{subsec:scaling_up}
%

All tests that have been discussed so far were conducted using a reduced dataset consisting of a single underlying (see Section \ref{subsec:datasets}). To assess the impact of increased problem dimensionality on the runtime of our model, we attempted to solve the mean--variance problem formulated by Eq. \eqref{eq:markowitz_problem} using the QUBO formulation on the full dataset. This setting increased the dimension of the solution vector $\mathbf{x}$ from $882$ to $11\,676$.

One of the main reasons for choosing \Gurobi{} was
its high efficiency when using the reduced dataset, 
consistently finding constrained solutions within a few seconds.

However, even \Gurobi{} struggles to find a solution to constrained instances of problems using the complete dataset. In particular, on Machine~B (32~GB of RAM), the MIQP solver ran out of memory without producing any solution, while \Gurobi \_Q was able to find a suboptimal solution. On Machine~A (57~GB of RAM), \Gurobi{} MIQP solver again ran out of memory before the optimal solution to the constrained problem was found, after approximately $15$ minutes, returning a high-quality solution. It is important to note that the pre-solving phase requires about $10$ minutes, thus imposing a lower bound on the total runtime.
Alternatively, \Gurobi\_Q with Lagrangian relaxation is capable of providing a good quality solution, although not as good as that found by \Gurobi{} in the constrained formulation, in $2.8$ minutes and could be run longer to refine such a solution without encountering out-of-memory issues.

This evidence suggests that, as the problem dimension grows, the QUBO approach may scale better than the constrained one. The constrained formulation becomes increasingly limited by \Gurobi{} memory usage and pre-solve time, whereas the QUBO route remains practically viable.

Notably, even if Simulated Bifurcation failed to find a feasible solution for the full dataset, it proved to be a memory-efficient approach compared to \Gurobi{}-based methods. This suggests that this QUBO formulation could further benefit from more reliable or better tunable quantum-inspired solvers, where improved controllability and robustness may enhance scalability and convergence on large-scale instances. In fact, in the implementation of the Simulated Bifurcation we used, the relationship between hyperparameter settings and solution quality is not fully transparent, making systematic tuning challenging and often leading to inconsistent performance across instances.

\section{Conclusion and future work}
\label{sec:conclusion}

We presented a framework for portfolio
optimization that combines exponential tilting of
empirical scenarios with quadratic surrogate models compatible
with QUBO solvers. The method constructs tilted moments via
Esscher reweighting of base-point P\&L, yielding a
tractable QUBO representation of a locally approximated CARA
certainty-equivalent objective that enables the use of
classical and quantum-inspired optimization tools. Varying the
tilt parameter induces controlled changes in 
moments, allowing the optimization to explore different 
regimes without altering the underlying scenario set.
Experiments on a realistic trading portfolio showed that
varying the tilt parameter produces portfolios with meaningfully
different risk--return profiles. For the QUBO formulation, Lagrangian relaxation proved
effective in encoding sensitivity inequality constraints without expanding
the binary search space. A preliminary scaling test (from
$882$ to $11\,676$ variables) suggested that the QUBO route
remains practically viable in larger instances, while
constrained MIQP encountered memory limitations.

Future work includes structured multi-tilt constructions, data-driven tilt selection targeting out-of-sample performance, and execution on quantum hardware to assess whether the observed QUBO scalability translates into practical benefit relative to classical and quantum-inspired methods.


\printbibliography

\newpage



\clearpage 
\onecolumn 
\appendices 

\numberwithin{equation}{section}

\makeatletter
\renewcommand{\thesubsectiondis}{\thesection.\arabic{subsection}.}
\renewcommand{\thesubsubsectiondis}{\thesection.\arabic{subsection}.\arabic{subsubsection}.}
\makeatother

\renewcommand{\thesubsection}{\thesection.\arabic{subsection}}
\renewcommand{\thesubsubsection}{\thesection.\arabic{subsection}.\arabic{subsubsection}}

\section{Sample Average Approximation and Esscher tilting}
\label{app:setup-basic-identities}

\subsection{Sample Average Approximation (SAA) on a finite scenario set}
\label{app:E:SAA}

Fix a base-point $\bar{\mathbf{x}}$ and a tilt $t$. We distinguish between two levels.
At the \emph{theoretical} level, scenarios are viewed as realizations of an underlying
probability law $\mathbb{P}$.
Under this law, the base-point P\&L $z(\bp)$ is a function of scenarios and can be seen as a scalar random variable, and tilted moments are expectations under the Esscher-tilted measure $\pi(t|\bp)$ obtained from $\mathbb{P}$.
We refer to these ideal quantities as \emph{population} quantities. 

At the \emph{empirical} level, the law $\mathbb{P}$ is unknown: we observe only a finite
scenario set. We write
$\pls{s}{\bp}$ for the realization of $z(\bp)$ in scenario $s$, and collect these
$\nS$ values in the vector $\pl{\bp} = (\pls{1}{\bp},\dots,\pls{\nS}{\bp})$. Thus,
$\pl{\bp}$ is the sample of realizations of the single random variable $z(\bp)$ across
the scenario set; only this sample is observable.
The sample average approximation
(SAA) replaces each population expectation by a weighted average over these scenarios.
Throughout, $\bar{\mathbf{x}}$ is held fixed (Appendix~\ref{app:base-point_choice}), so
$\pl{\bp}$ is a fixed scenario vector and the only source of randomness is the scenario
sampling.

At the theoretical level, the Esscher tilt reweights $\mathbb{P}$ by the base-point
P\&L.
For any scenario quantity $\varphi$, the corresponding population tilted
expectation is
\begin{equation*}
{
\E_{{\pi}(t|\bp)}[\varphi]
=
\frac{\E_{\mathbb{P}}\bigl[\varphi\,e^{t\, z(\bp)}\bigr]}{\E_{\mathbb{P}}\bigl[e^{t\, z(\bp)}\bigr]} .
}
\end{equation*}
Since $\mathbb{P}$ is unknown, the SAA replaces the population expectations above by
averages over the observed scenarios: the empirical mean serves as a surrogate for the
unknown expectation under $\mathbb{P}$.

Let $w_s>0$, with $\sum_{s=1}^\nS w_s=1$, be the base weights on the scenario set. The empirical (Esscher/softmax) weights are the ones defined in Eq.~\eqref{eq:esscher-weights}, and the empirical tilted expectation is the weighted average
\begin{equation*}
{
\E_{\boldsymbol{\pi}(t|\bp)}[\boldsymbol\varphi]
=
\sum_{s=1}^\nS \pi_s(t|\bp)\,\varphi_s .
}
\end{equation*}
Hence, tilted moments are computed by direct weighted averaging.

Assume the scenarios are independent and identically distributed draws from $\mathbb{P}$ with $w_s=1/\nS$, and $\E_{\mathbb{P}}[e^{t z(\bp)}]<\infty$. By the strong law of large numbers applied to numerator and denominator, for any integrable $\varphi$ the empirical weighted average converges to the population tilted expectation,
\begin{equation*}
{
\sum_{s=1}^\nS \pi_s(t|\bp)\,\varphi_s
\;\xrightarrow[\nS\to\infty]{a.s.}\;
\frac{\E_{\mathbb{P}}\bigl[\varphi\,e^{t\, z(\bp)}\bigr]}{\E_{\mathbb{P}}\bigl[e^{t\, z(\bp)}\bigr]} .
}
\end{equation*}
Thus the empirical tilted moments are consistent estimators of the population tilted moments at each
fixed $(t,\bar{\mathbf{x}})$.

If base weights $w_s$ are non-uniform, all formulas above remain valid.


\subsection{Variational/KL viewpoint (score--entropy trade-off) -- characterization of the Esscher weights.}

The same weights arise as the unique optimizer of a strictly concave score--entropy trade-off: increasing the expected baseline score while staying close to the base measure. Concretely,
\[
\boldsymbol{\pi}(t|\bp)
=\arg\max_{p\in\Delta_\nS}\Bigl\{\, t\sum_{s=1}^\nS p_s\,\pls{s}{\bp}
-\mathrm{KL}(p\|\mathbf{w})\Bigr\},
\]
where $\Delta_\nS$ is the probability simplex and
\[
\mathrm{KL}(p\|\mathbf{w})
=
\sum_{s=1}^\nS p_s \log\!\left(\frac{p_s}{w_s}\right)
\]
is the Kullback--Leibler divergence (relative entropy) of $p$
with respect to the base measure $\mathbf{w}$.
It is non-negative and vanishes if and only if $p=\mathbf{w}$.
The KL divergence is not a metric -- it is generally not symmetric and does not satisfy the triangle inequality -- but, being non-negative and zero only at $p=\mathbf{w}$, it quantifies the discrepancy between the candidate measure $p$ and the base measure $\mathbf{w}$.
Hence, the optimization problem balances two competing effects:
increasing the expected stressed score while remaining close to
the original scenario measure. When $\mathbf{w}$ is uniform, $-\mathrm{KL}(p\|\mathbf{w})$ differs from the Shannon entropy $H(p)$ only by an additive
constant, so the objective is equivalently a score--entropy trade-off. The associated value function is
the log-sum-exp (a smooth maximum), explaining the increasing concentration of $\boldsymbol{\pi}(t|\bp)$ as $|t|$ grows.

This variational characterization is a finite-scenario instance of the
Donsker--Varadhan variational formula for relative entropy~\cite{dembo2010large}.
In the portfolio context, the same relative-entropy reweighting principle is closely
related to the entropy-pooling construction of~\cite{meucci2008entropypooling},
where posterior scenario probabilities are obtained by minimal relative-entropy
deformation of a base measure under views.

In conclusion, the softmax definition in Eq.~\eqref{eq:esscher-weights} specifies how the tilted weights are computed; the variational characterization above clarifies what they represent. The tilted measure can be read as the solution of a constrained reweighting problem: among all measures on the scenario set, $\boldsymbol{\pi}(t|\bp)$ is the one that attains a given level of expected base-point score with the smallest departure -- in relative-entropy terms -- from the base measure $\mathbf{w}$. This gives the tilt a precise meaning: $t$ is the exchange rate between expected score and fidelity to $\mathbf{w}$, and the monotone concentration of the weights as $|t|$ grows reflects the score term progressively dominating the entropic penalty.

\section{base-point selection}\label{app:recentering}

\subsection{Choosing the base-point $\bar{\mathbf{x}}$}\label{app:base-point_choice}

The exponential tilting mechanism requires a base-point $\bar{\mathbf{x}}$, for which several choices are possible. In our experiments, we choose to set  $\bar{\mathbf{x}}$ as the solution of the continuous relaxation of a standard mean--variance Markowitz problem computed under the base scenario weights, imposing the same constraints as in the downstream step. This problem is convex and yields a robust baseline from which to perform reweighting, as its solution typically lies in the neighborhood of high-quality portfolios.
Other valid choices are, for example, the initial portfolio or a feasible point found using a greedy or multi-start heuristic.

In all cases, once $\bar{\mathbf{x}}$ is fixed, our pipeline constructs $\boldsymbol{\pi}(t|\bp)$, then
$\bar{\mathbf{b}}_{t,\bar{\mathbf{x}}},\bar{\mathbf{Q}}_{t,\bar{\mathbf{x}}}$, and solves the resulting quadratic problem. It is worth noting that the procedure can be iterated by using the solution of the tilted problem as a new base-point, thereby further refining the portfolio obtained in the previous iteration.

\subsection{Markowitz base-point properties}
\label{app:F:psd}

For completeness, let
\[
{
f_{\text{abs}}(\mathbf{x})
:=
\frac{1}{2}\mathbf{x}^{\top}\boldsymbol{\Sigma}_{GG}\mathbf{x}
+\boldsymbol{\Sigma}_{G\initp}^{\top}\mathbf{x}
-\boldsymbol{\mu}_{G}^{\top}\mathbf{x}\,
}
\]
be the non-constant part of the Markowitz objective in Eq.~\eqref{eq:markowitz_obj}.

The matrix $\boldsymbol{\Sigma}_{GG}$ is positive semidefinite (PSD). Indeed, recall
from Eq.~\eqref{eq:w_cov} that, under the uniform measure $\mathbf{m}$,
\[
\boldsymbol{\Sigma}_{GG}
=
\m{G}^{\top}\bigl(\mathrm{diag}(\mathbf{m}) - \mathbf{m}\mathbf{m}^{\top}\bigr)\m{G},
\]
where $\m{G}\in\mathbb{R}^{\nS\times d}$. Let
$\m{C} := \mathrm{diag}(\mathbf{m}) - \mathbf{m}\mathbf{m}^{\top}$. For any
$\mathbf{u}\in\mathbb{R}^{\nS}$, writing $\bar u := \sum_{s=1}^{\nS} m_s u_s$,
\[
\mathbf{u}^{\top}\m{C}\,\mathbf{u}
=
\sum_{s=1}^{\nS} m_s u_s^{2} - \Bigl(\sum_{s=1}^{\nS} m_s u_s\Bigr)^{2}
=
\sum_{s=1}^{\nS} m_s\,(u_s-\bar u)^{2}
\;\ge\;0,
\]
so $\m{C}$ is PSD (this quantity is the variance of the scenario vector $\mathbf{u}$ under $\mathbf{m}$).
Hence, for any $\mathbf{v}\in\mathbb{R}^{d}$, setting $\mathbf{u}=\m{G}\mathbf{v}$,
\[
\mathbf{v}^{\top}\boldsymbol{\Sigma}_{GG}\mathbf{v}
=
(\m{G}\mathbf{v})^{\top}\m{C}\,(\m{G}\mathbf{v})
\;\ge\;0,
\]
which proves that $\boldsymbol{\Sigma}_{GG}$ is PSD.

If the initial portfolio P\&L is uncorrelated with the candidate instruments in the empirical
scenario set, i.e., $\boldsymbol{\Sigma}_{\m{G}\initp}=\mathbf{0}$, then the objective reduces
to the form
\[
f_{\text{inc}}(\mathbf{x})
\;:=\;
\frac{1}{2}\mathbf{x}^{\top}\boldsymbol{\Sigma}_{\m{G}\m{G}}\mathbf{x}
-\boldsymbol{\mu}_{\m{G}}^{\top}\mathbf{x},
\]
with the same PSD curvature matrix $\boldsymbol{\Sigma}_{\m{G}\m{G}}$.

Consequently, both $f_{\text{abs}}(\mathbf{x})$ and $f_{\text{inc}}(\mathbf{x})$ are convex
quadratic objectives on $\mathbb{R}^{d}$, and remain convex under standard convex constraints
(e.g., box constraints $0\le x_u\le 1$ and linear budget constraints).

\section{Inequality constraints: Slack variables and Lagrangian Relaxation}\label{app:lagrabgian}

\subsection{Slack variables experimental inefficiency}
\label{app:slack}

The slack variable approach to integrating the Greeks constraints into the QUBO formulation 
proved underperforming and difficult to generalize. In this method, the solution space is 
extended by introducing auxiliary binary variables to represent the feasible range of 
solutions satisfying the sensitivity constraints. The slack variable method is designed to 
yield a zero-valued penalty (or near-zero, depending on the discretization of the variable) 
whenever the solution complies with the Greeks constraints, and a large penalty when they are 
violated. This weight of the penalty is not fixed across instances of the optimization problem, but is rather calibrated relative to the other penalty terms and the scale of the objective function. Specifically, we weighted the penalty in Equation \ref{eq:slack_penalty} with

$$
w_{\mathcal{G}} = w_{\mathrm{greek}}\frac{(w_{\text{EOS}}+w_{\text{UEI}})}{2C^2_\mathcal{G}},
$$

where $w_{\text{EOS}}$ and $w_{\text{UEI}}$ are the weights associated respectively with the EOS composition constraint and with the UEI quantities constraint. Instead, $w_{\mathrm{greek}}$ is a penalty value parameter that we tried to tune to allow each problem instance to reach feasible solutions concerning all the Greeks constraints.

\begin{figure}[h]
    \centering

    \begin{subfigure}{\linewidth}
        \centering
        \label{fig:gurobi_block}
        \begin{subfigure}[b]{0.3\linewidth}
            \includegraphics[width=\linewidth]{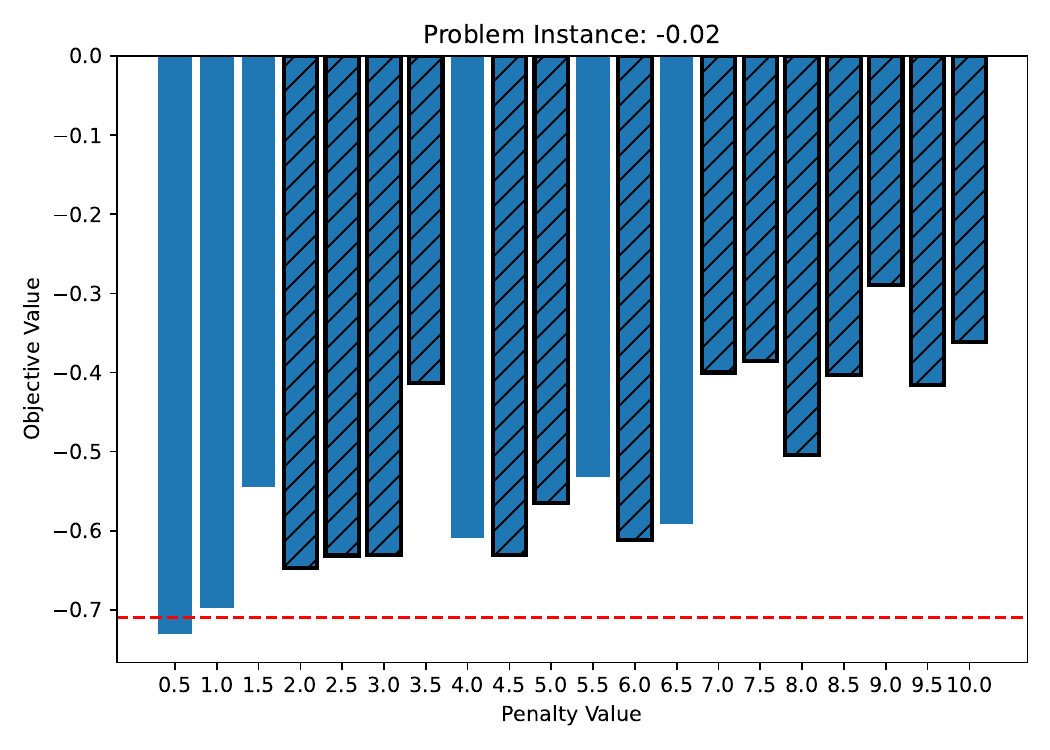}
        \end{subfigure}
        \hfill
        \begin{subfigure}[b]{0.3\linewidth}
            \includegraphics[width=\linewidth]{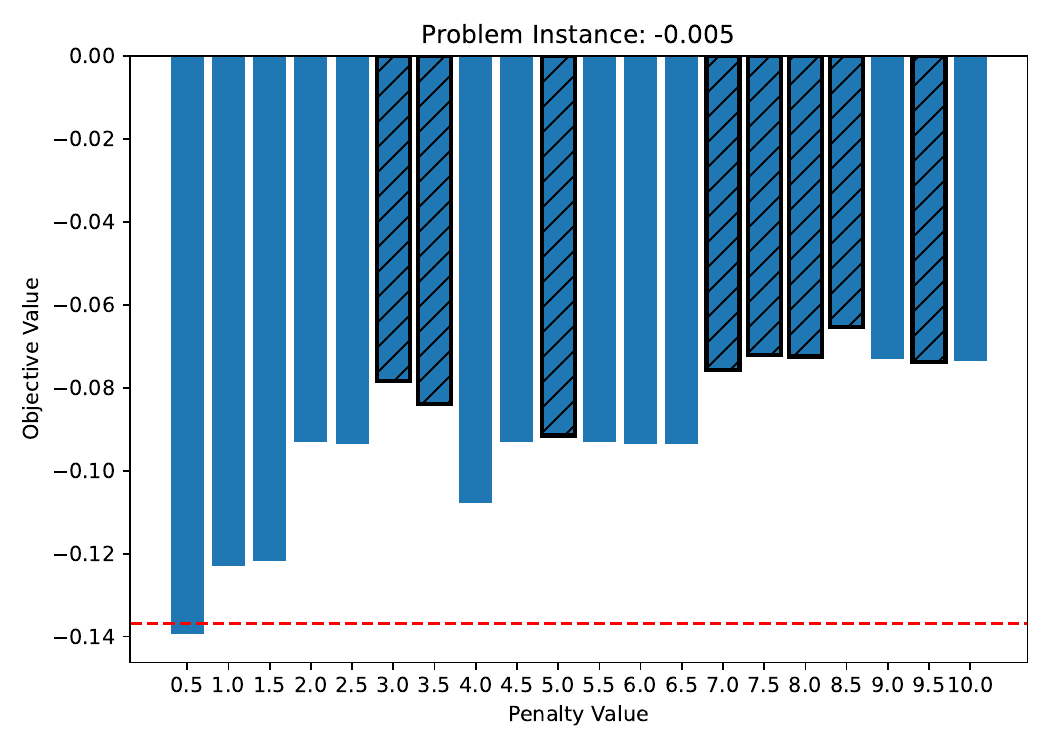}
        \end{subfigure}
        \hfill
        \begin{subfigure}[b]{0.3\linewidth}
            \includegraphics[width=\linewidth]{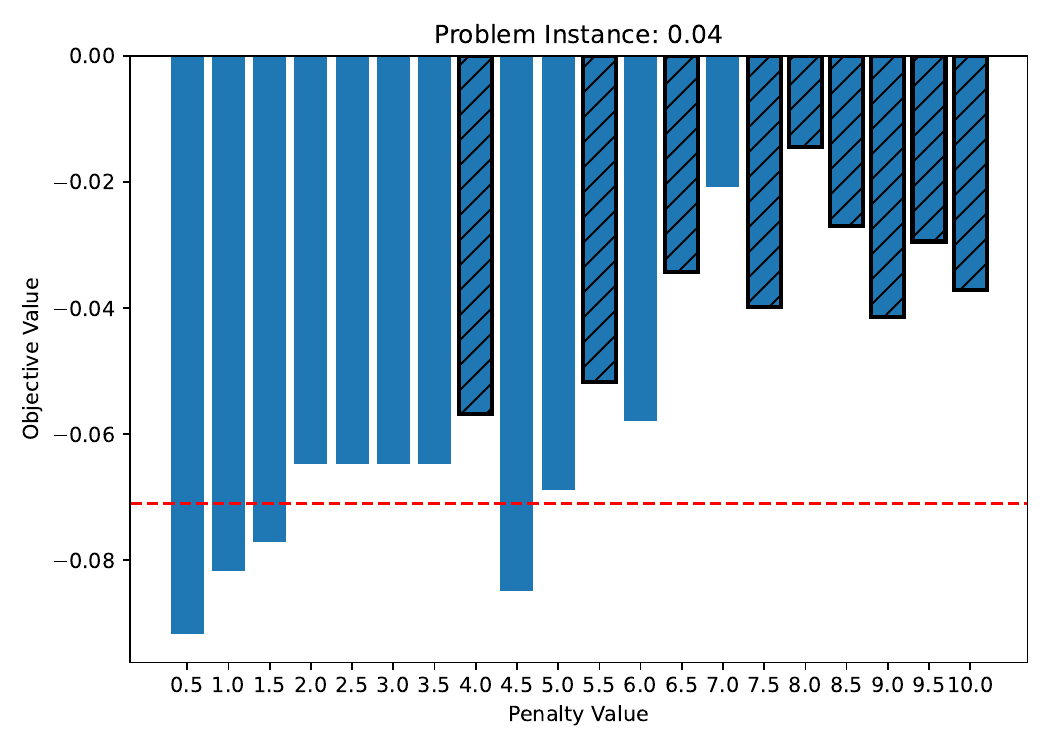}
        \end{subfigure}

        \vspace{0.4em}

        \begin{subfigure}[b]{0.3\linewidth}
            \includegraphics[width=\linewidth]{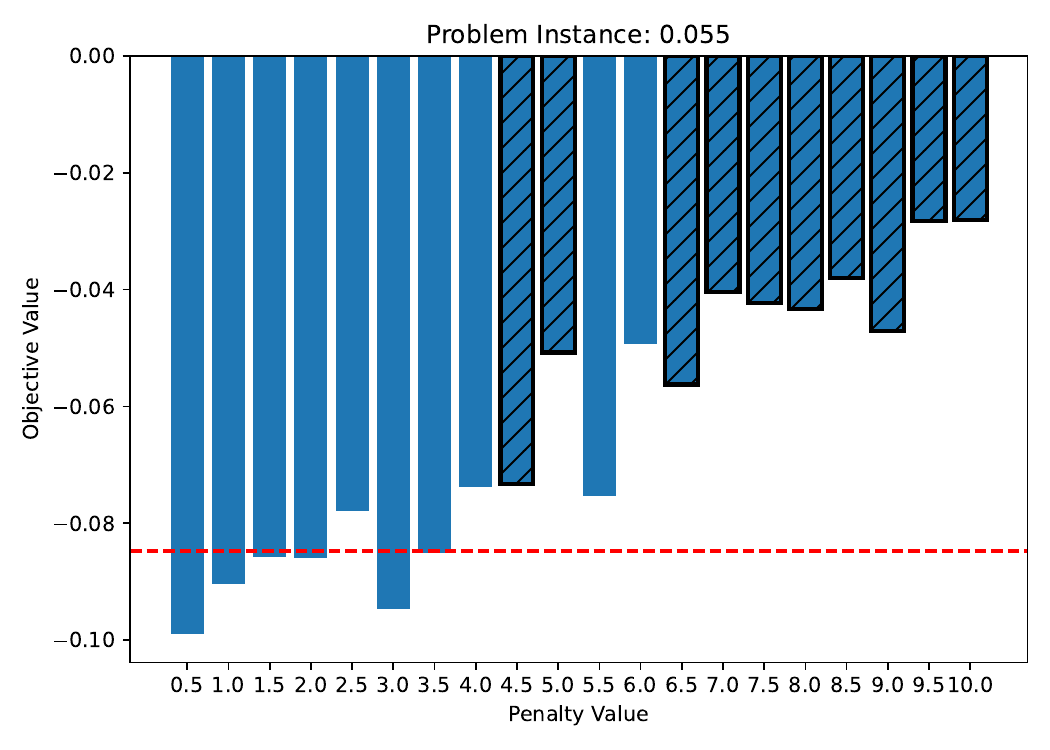}
        \end{subfigure}
        \hfill
        \begin{subfigure}[b]{0.3\linewidth}
            \includegraphics[width=\linewidth]{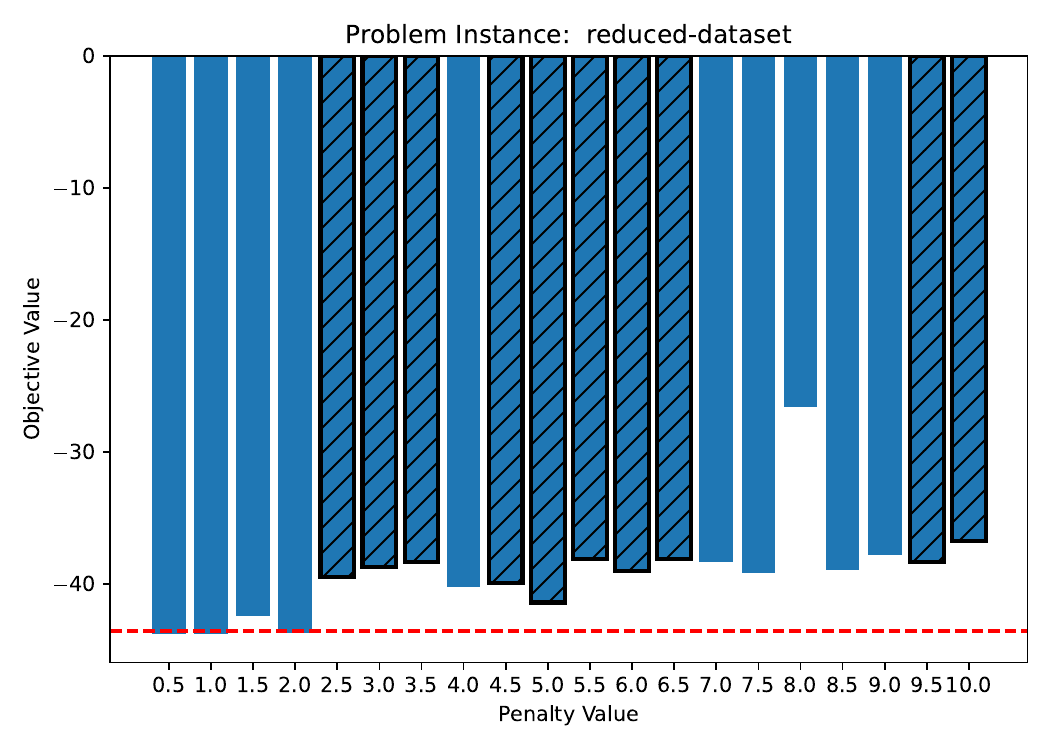}
        \end{subfigure}
        \hfill
        \begin{subfigure}[b]{0.3\linewidth}
            \includegraphics[width=\linewidth]{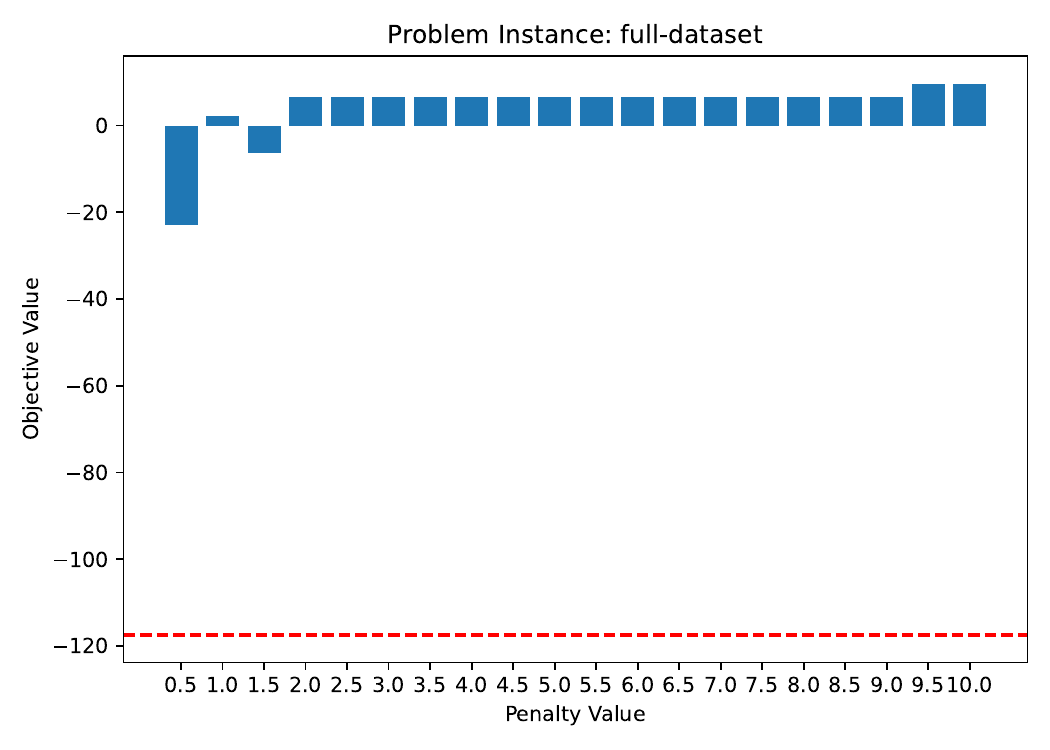}
        \end{subfigure}

        \caption{\Gurobi\_Q results.}
        \label{fig:gurobi_block}
    \end{subfigure}

    \vspace{1em}

    \begin{subfigure}{\linewidth}
        \centering
        \label{fig:simbif_block}
        \begin{subfigure}[b]{0.3\linewidth}
            \includegraphics[width=\linewidth]{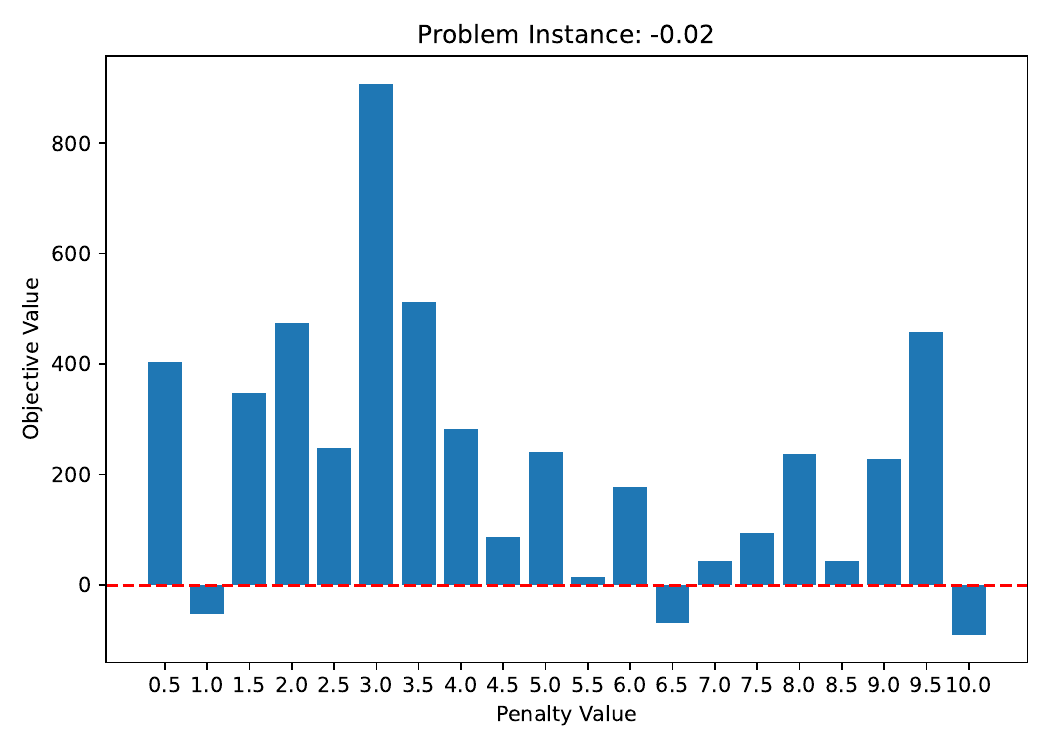}
        \end{subfigure}
        \hfill
        \begin{subfigure}[b]{0.3\linewidth}
            \includegraphics[width=\linewidth]{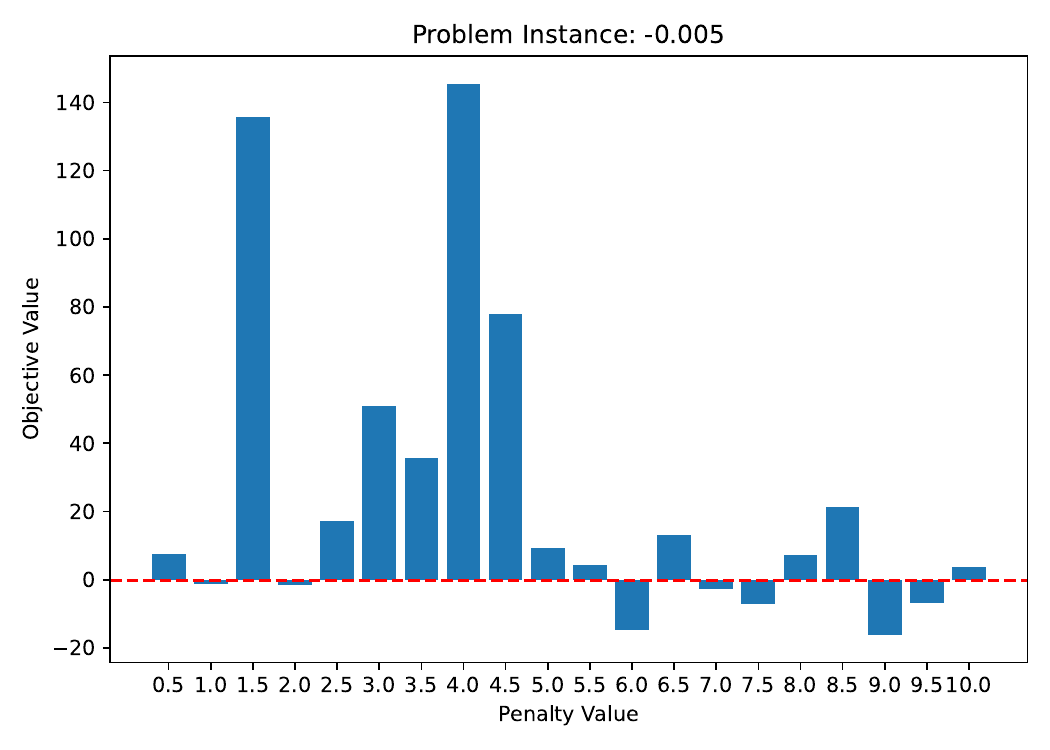}
        \end{subfigure}
        \hfill
        \begin{subfigure}[b]{0.3\linewidth}
            \includegraphics[width=\linewidth]{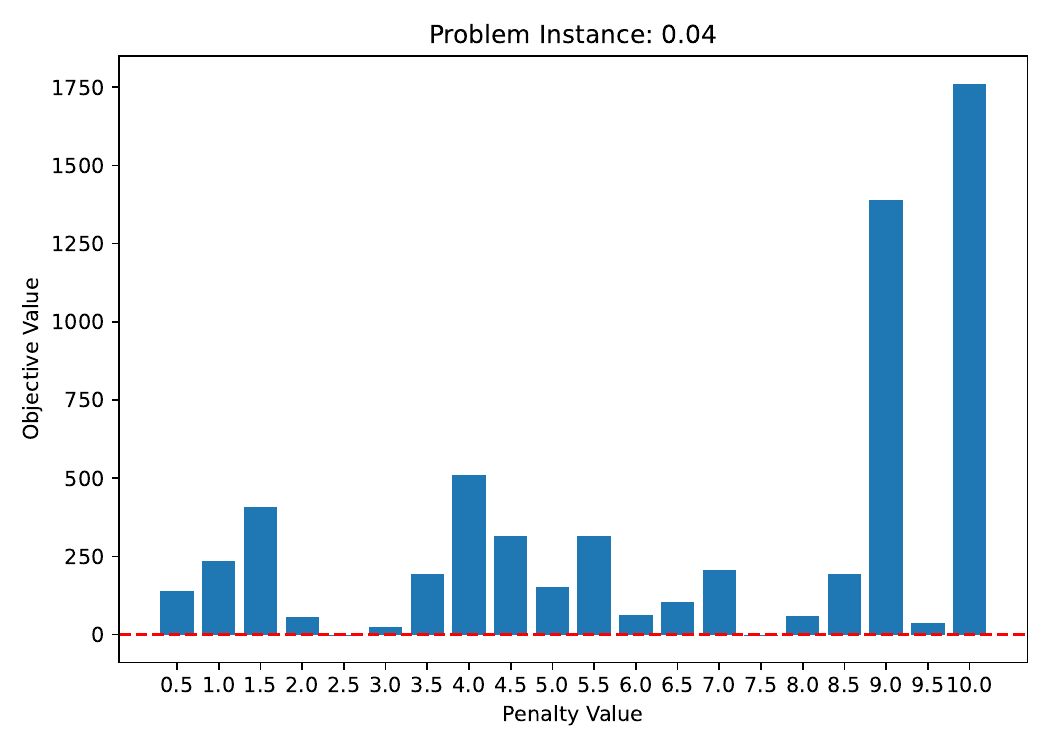}
        \end{subfigure}

        \vspace{0.4em}

        \begin{subfigure}[b]{0.3\linewidth}
            \includegraphics[width=\linewidth]{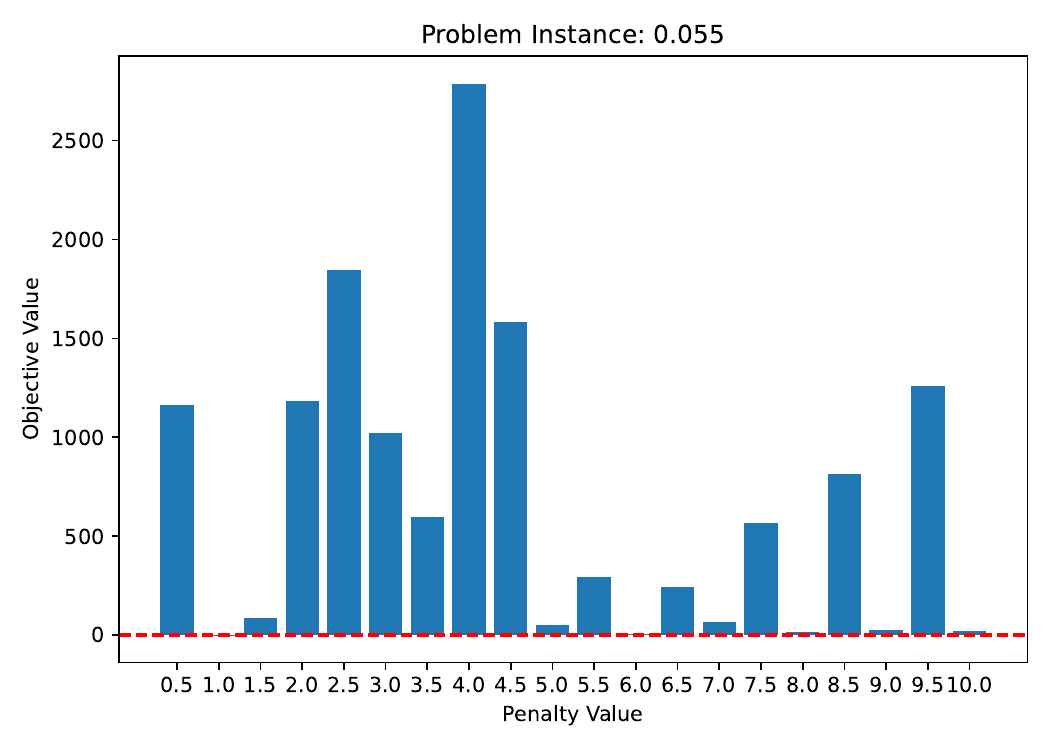}
        \end{subfigure}
        \hfill
        \begin{subfigure}[b]{0.3\linewidth}
            \includegraphics[width=\linewidth]{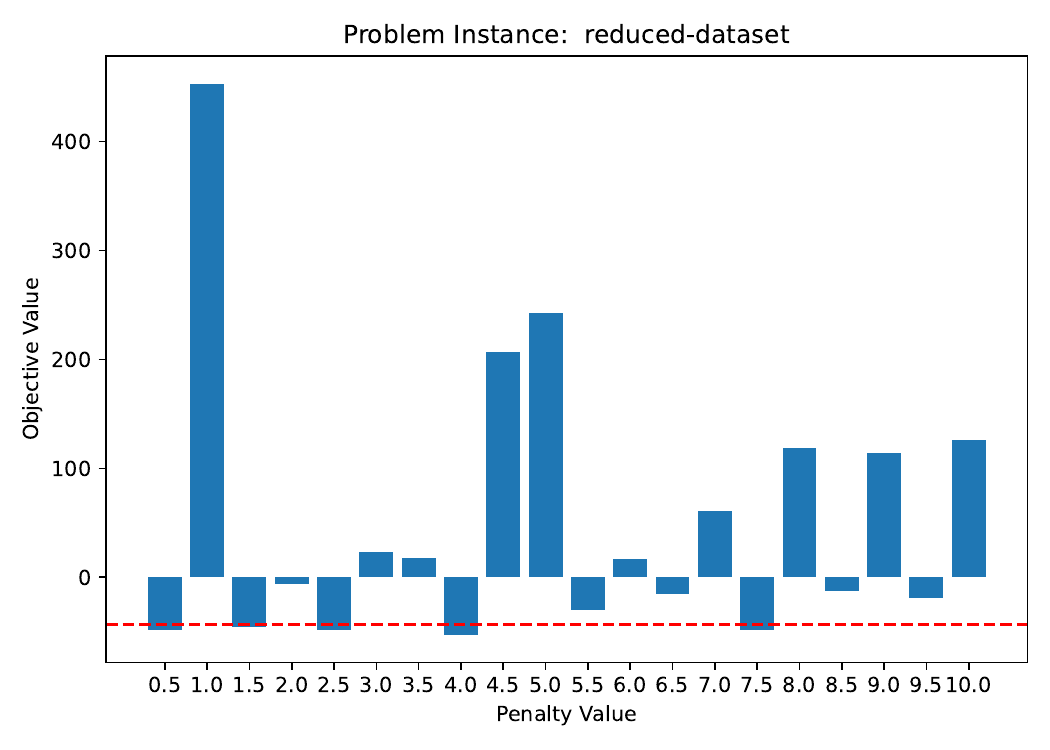}
        \end{subfigure}
        \hfill
        \begin{subfigure}[b]{0.3\linewidth}
            \includegraphics[width=\linewidth]{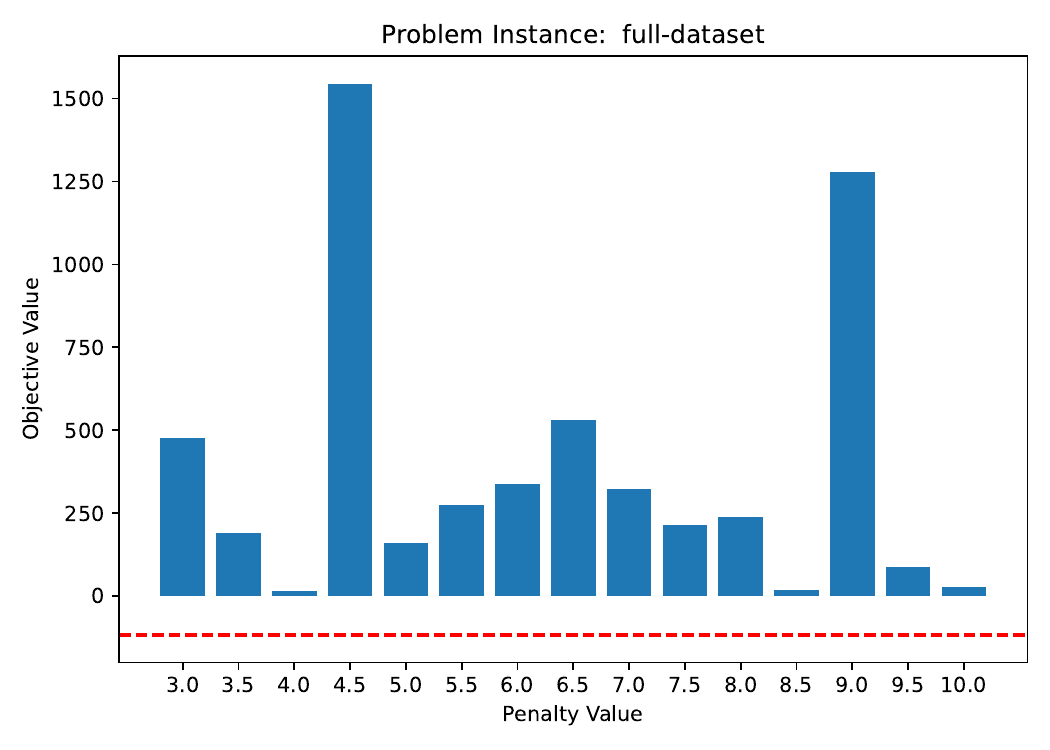}
        \end{subfigure}

        \caption{Simulated Bifurcation results. }
        \label{fig:sim_bif_block}
    \end{subfigure}

    \caption{Results for the slack variable method using \Gurobi\_Q (\subref{fig:gurobi_block}) and Simulated Bifurcation (\subref{fig:sim_bif_block}) solvers, varying penalty value $w_{\mathrm{greek}}$. Red dashed lines indicate the optimal MIQP reference value; hatched bars denote feasible solutions.}
    \label{fig:comparison_gurobi_simbif}
\end{figure}

Figure~\ref{fig:gurobi_block} reports the results obtained across different penalty values 
for the 6 instances considered in the QUBO formulation, using \Gurobi\_Q. This solver 
exhibits stable behavior: as seen in the plots, increasing the penalty value tends to 
increase the likelihood of obtaining a feasible solution. However, none of the feasible 
solutions found matches the optimum of the constrained problem identified by \Gurobi{}.
The Simulated Bifurcation solver, by contrast, exhibits more unpredictable behavior. As 
reported in Figure~\ref{fig:sim_bif_block}, it fails to find any feasible solution across 
all instances, despite being allotted up to 1 hour of runtime. These results were obtained with \texttt{mode='discrete'}, which is generally slower but more accurate than the ballistic variant, and \texttt{heated=True} that further improves solution quality at the cost of additional computation. We experimented also with the other solver configurations and, although this setting still did not yield feasible solutions, it was the only one to produce non-trivial, non-zero outputs.

\subsection{Lagrangian relaxation experimental convergence}
\label{app:lagr}
 
In principle, our Lagrangian relaxation-based methodology would provide a certificate of convergence to the optimum of the dual problem; however, neither \Gurobi\_Q nor Simulated Bifurcation guarantee optimality of the individual QUBO subproblems. We therefore adopted a practical stopping criterion: termination upon matching the optimum objective value of the constrained problem found by \Gurobi{}, or after a maximum of 200~iterations, whichever occurred first.
 
\begin{figure}[h]
    \centering
    \begin{subfigure}[b]{0.49\linewidth}
        \includegraphics[width=\linewidth]{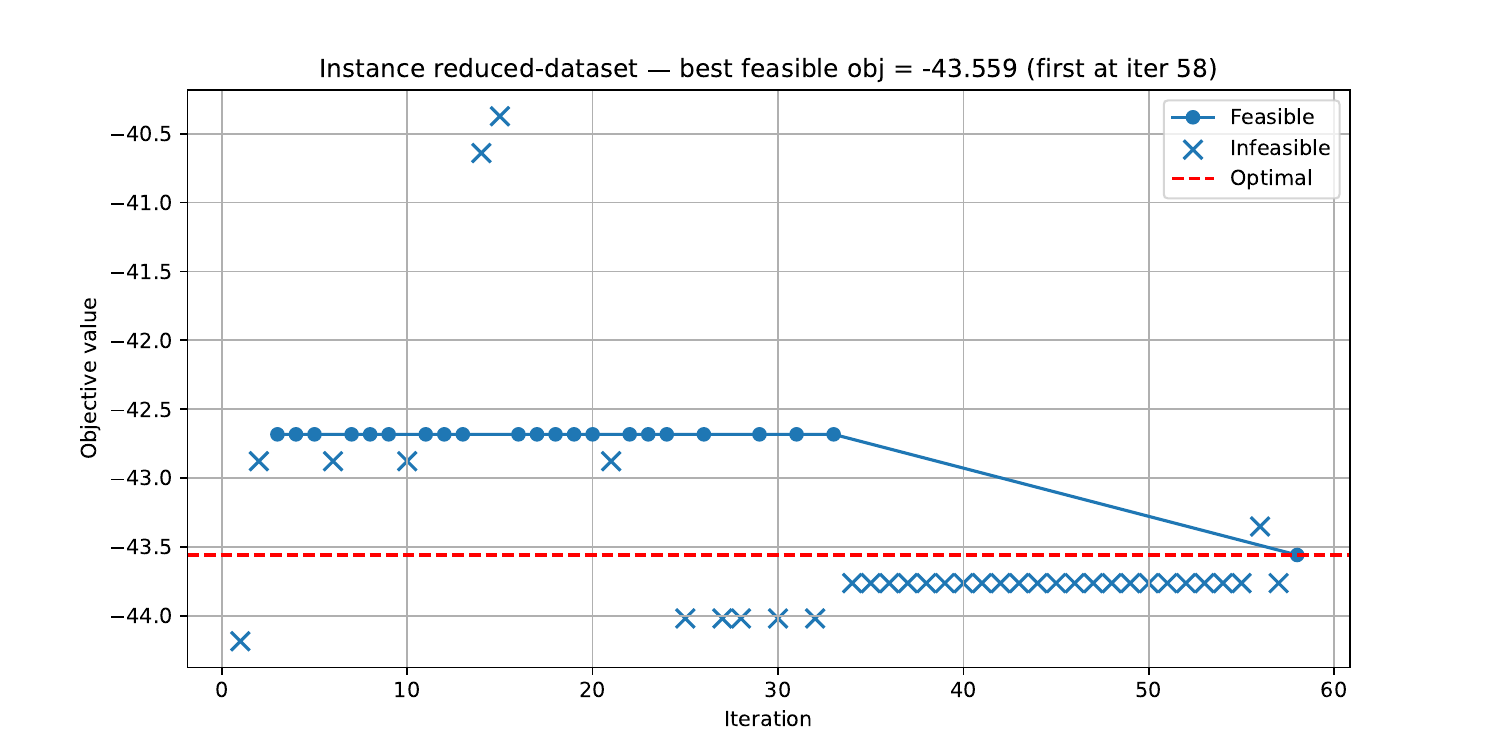}
    \end{subfigure}
    \hfill
    \begin{subfigure}[b]{0.49\linewidth}
        \includegraphics[width=\linewidth]{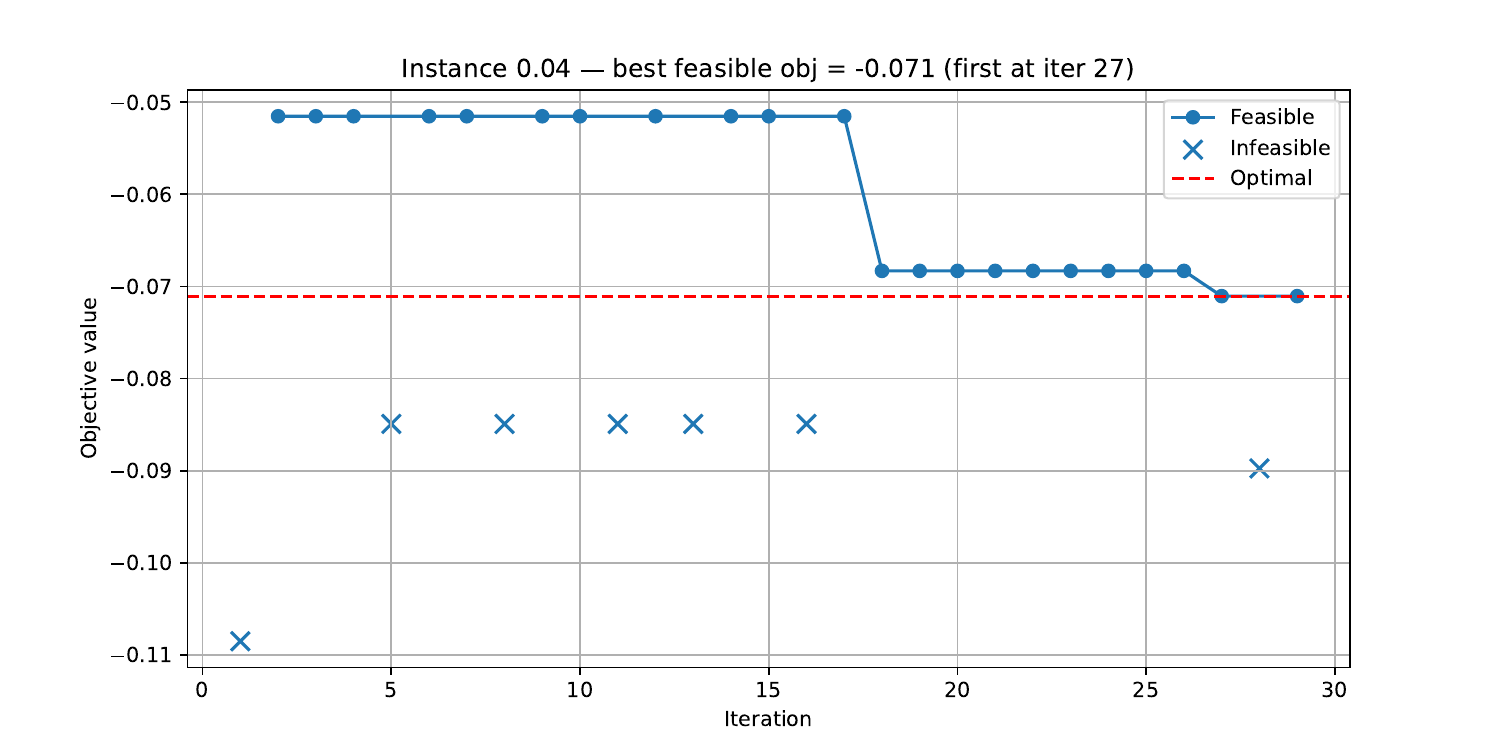}
    \end{subfigure}

    \vspace{0.5em} 

    \begin{subfigure}[b]{0.49\linewidth}
        \includegraphics[width=\linewidth]{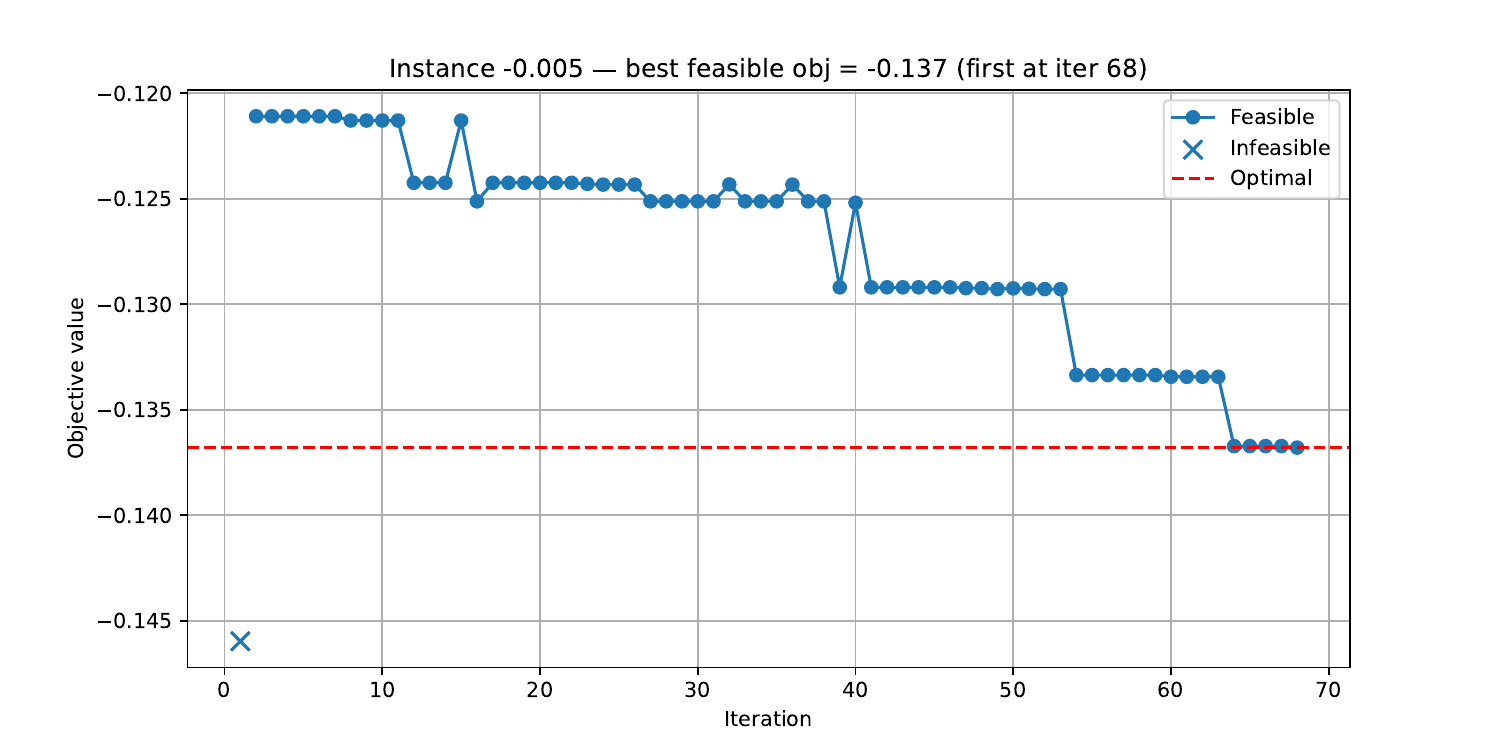}
    \end{subfigure}
    \hfill
    \begin{subfigure}[b]{0.49\linewidth}
        \includegraphics[width=\linewidth]{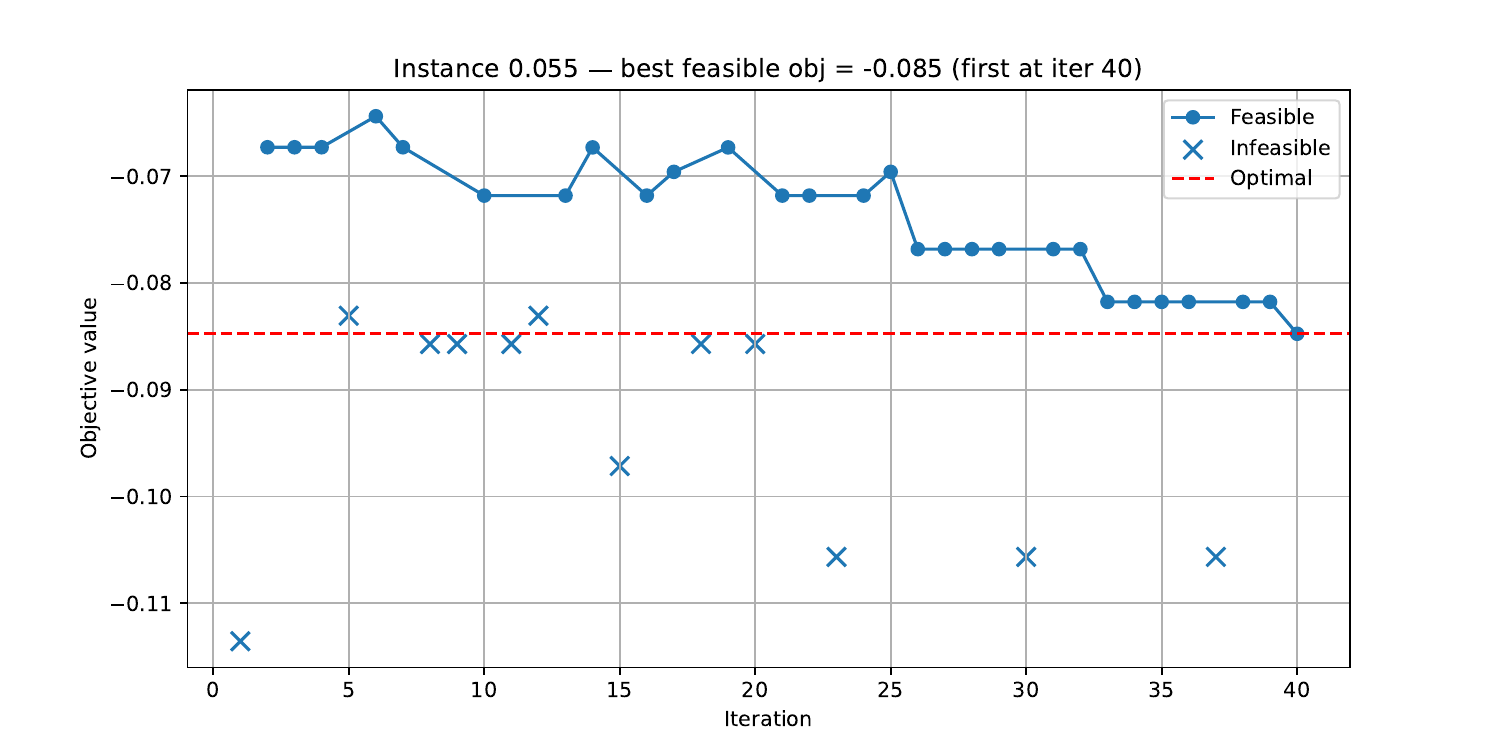}
    \end{subfigure}
    \hfill

    \caption{Lagrangian relaxation convergence for the four instances in which \Gurobi\_Q reached the \Gurobi{} MIQP optimal objective value (dashed line): reduced Markowitz dataset, $t=0.04$, $t=-0.005$, and $t=0.055$.}
    \label{fig:lr_opt}
\end{figure}

\begin{figure}[h]
    \centering
    \includegraphics[width=1\linewidth]{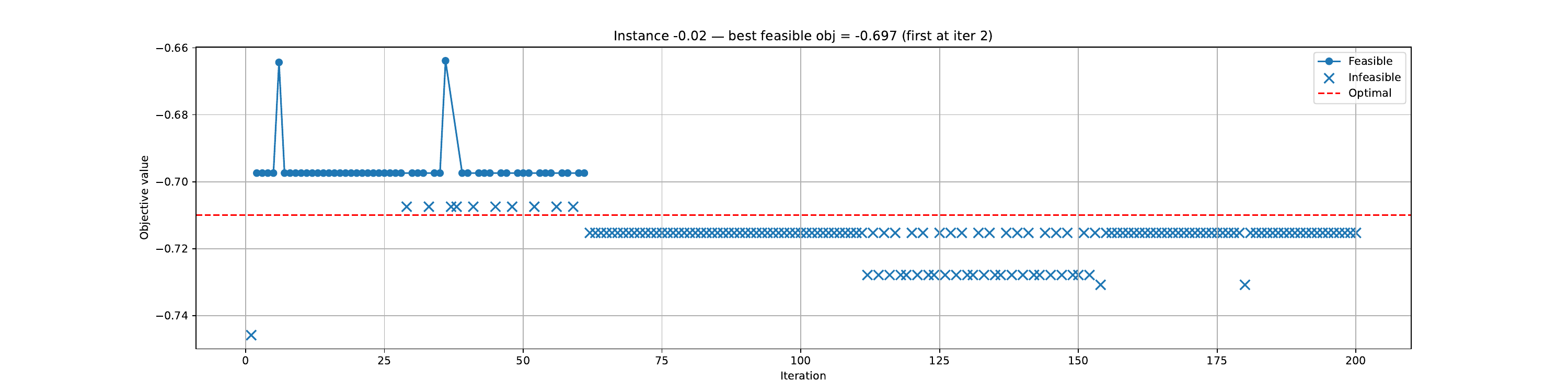}
    \caption{Lagrangian relaxation convergence for instance $t=-0.02$, using \Gurobi\_Q as QUBO solver. The optimal objective value of the constrained problem solved using \Gurobi{} (dashed line) is not reached; however, a solution with a $1.77\%$ optimality gap is already obtained at the second iteration.}
    \label{fig:gurobi002}
\end{figure}

\begin{figure}[h]
    \centering
    \includegraphics[width=1\linewidth]{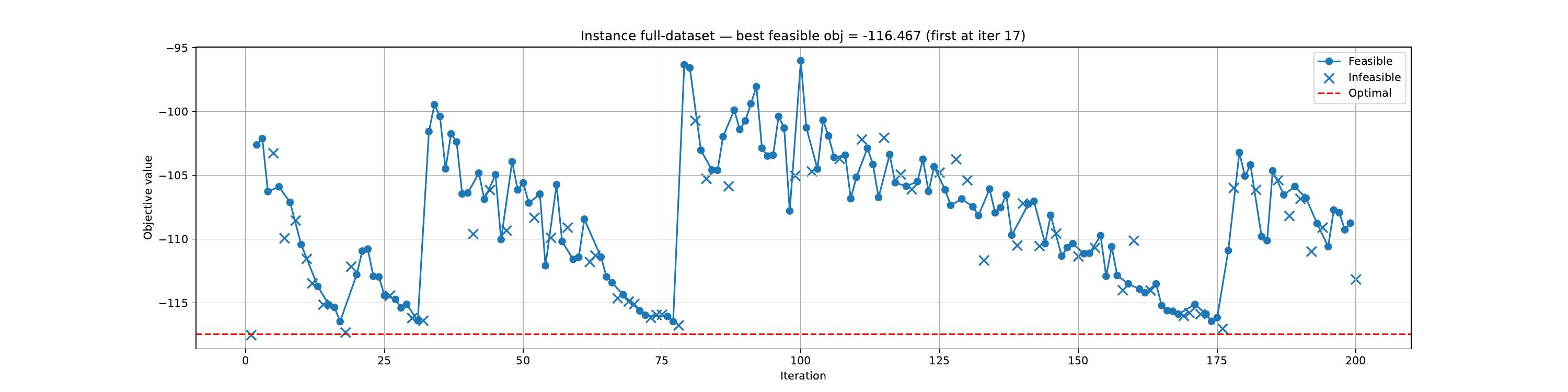}
    \caption{Lagrangian relaxation convergence for the Markowitz approach applied to the full-dataset, using \Gurobi\_Q as QUBO solver. The 
optimal objective value of the constrained problem solved using \Gurobi{} (dashed line) is not reached, however the best solution reached by \Gurobi\_Q was already found at iteration~17 in approximately 2.8~minutes.}
    \label{fig:gurobi_full}
\end{figure}

When using \Gurobi\_Q, four out of six instances converged to the optimal solution within 27--68~iterations, as illustrated in Figure~\ref{fig:lr_opt}. The two
remaining instances, $t = -0.02$ and the Markowitz approach applied to the full-dataset, did not reach the
target objective value. Nevertheless, both cases yielded noteworthy results: for $t = -0.02$, a high-quality solution with only a $1.77\%$ optimality gap was already obtained at the second iteration (Figure \ref{fig:gurobi002}); for the full-dataset instance, the best solution was found at iteration~17 (Figure \ref{fig:gurobi_full}) and, although it fell short of the \Gurobi{} MIQP optimum, it was produced in approximately 2.8~minutes, a time within which the MIQP solver had not yet completed its
presolving phase, and which caused an out-of-memory error on Machine~B.

\begin{figure}[h]
    \centering
    \includegraphics[width=1\linewidth]{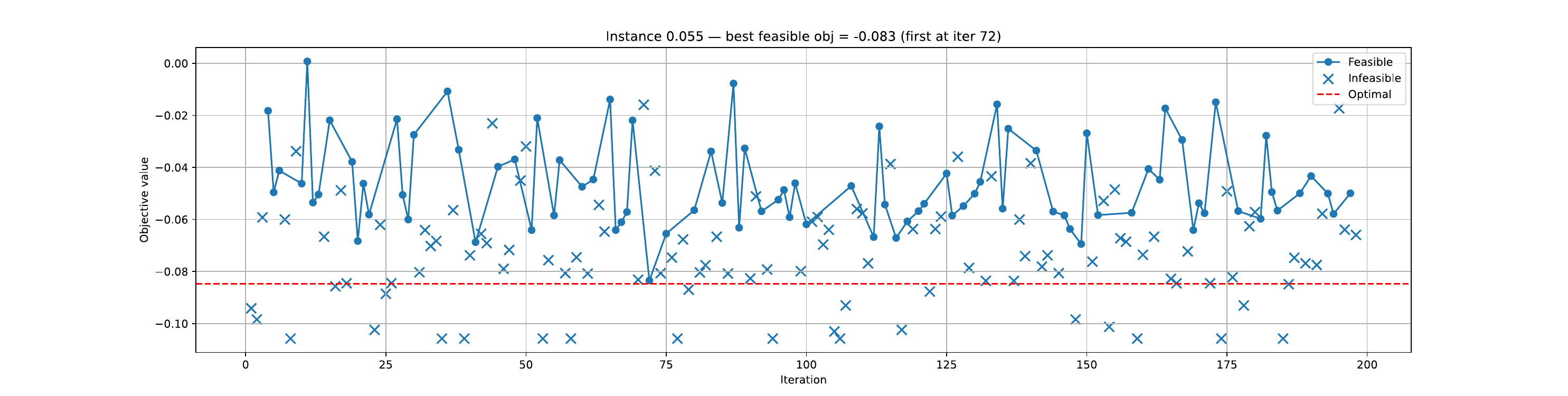}
    \caption{Lagrangian relaxation convergence for instance $t=0.055$ using Simulated
Bifurcation. No optimal solution is found within 200~iterations; the best solution obtained is within $1.5\%$ of the optimum in approximately 36 minutes.}
    \label{fig:sb_0055}
\end{figure}

\begin{figure}[h]
    \centering
    \begin{subfigure}[b]{0.49\linewidth}
        \includegraphics[width=\linewidth]{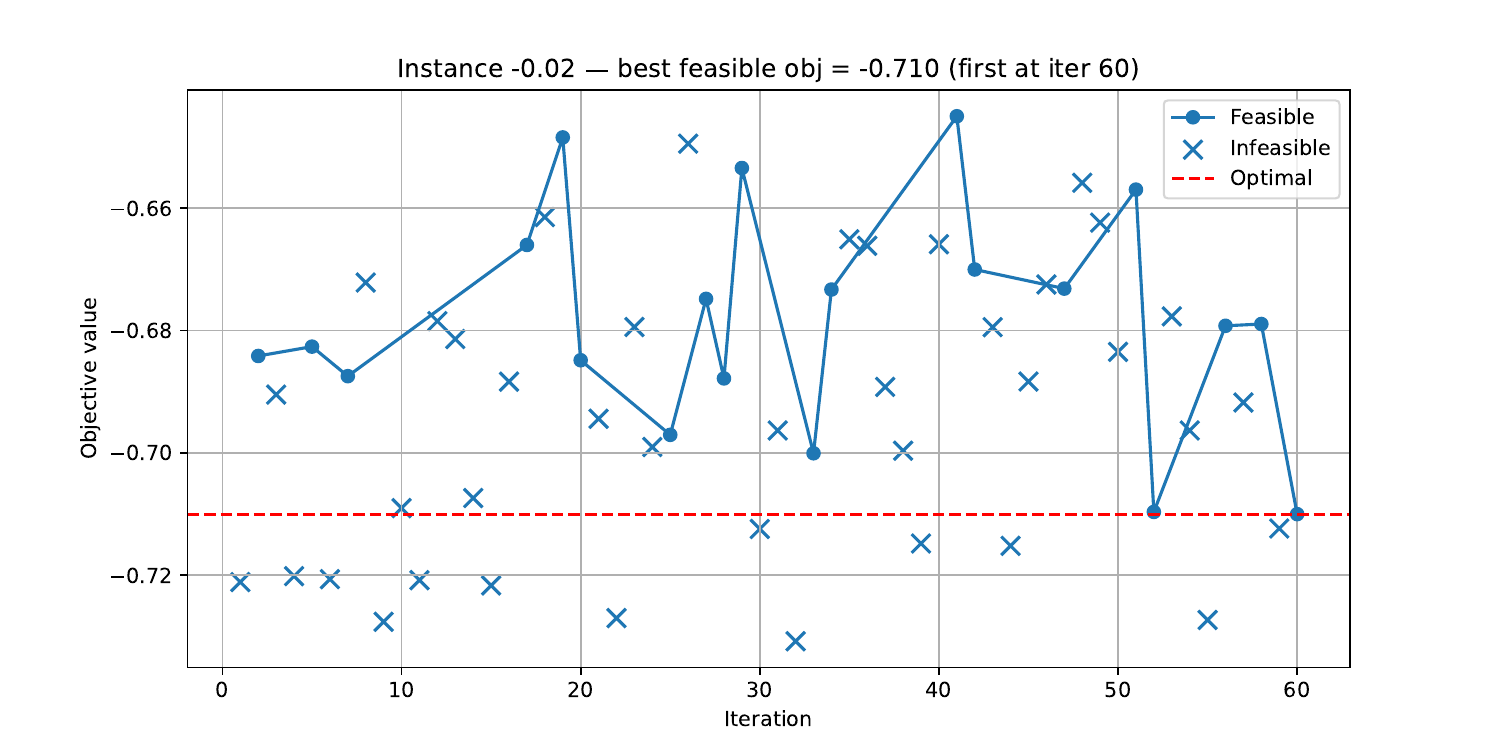}
        \caption{Instance $t=-0.02$: optimal solution found within 30 minutes}
        \label{fig:sb_002}
    \end{subfigure}
    \hfill
    \begin{subfigure}[b]{0.49\linewidth}
        \includegraphics[width=\linewidth]{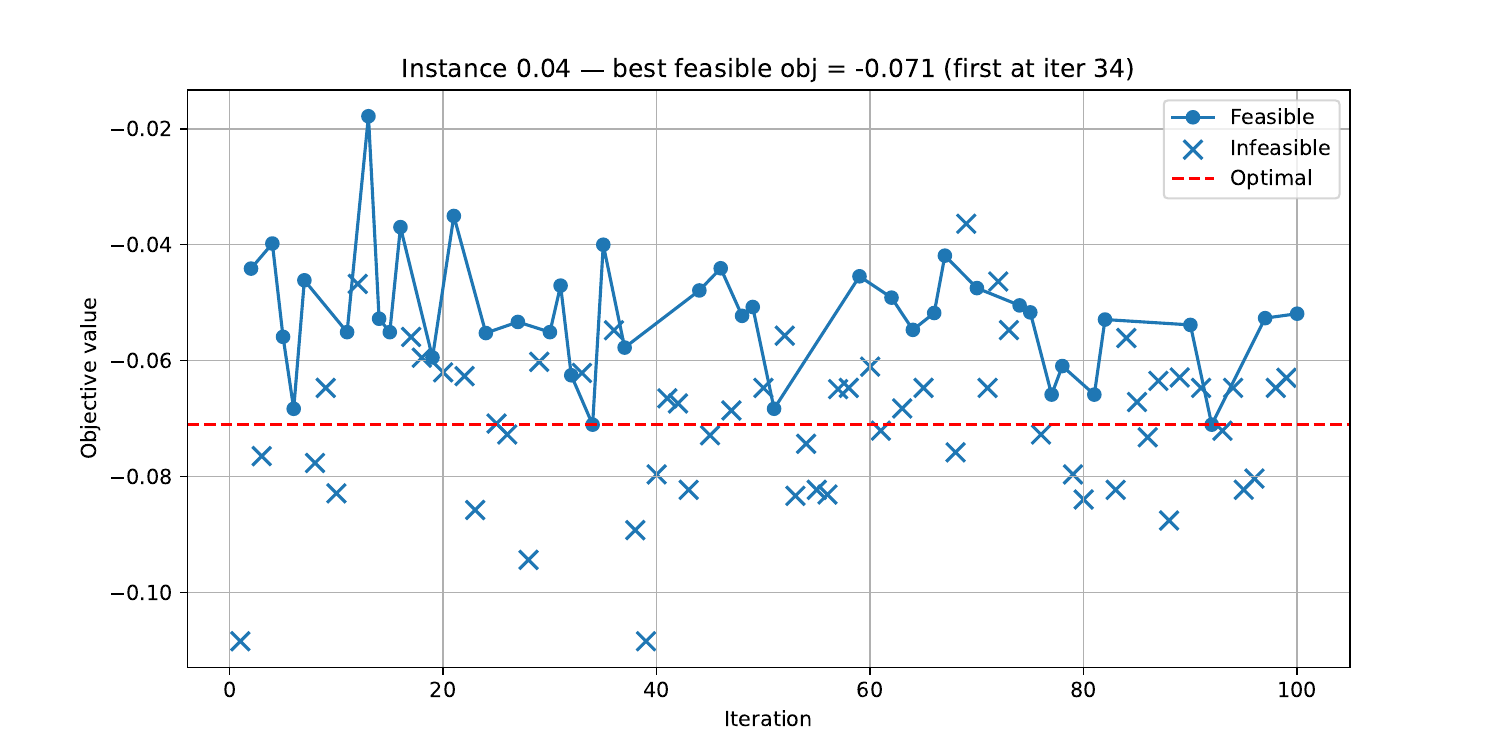}
        \caption{Instance $t=0.04$: optimal solution found within 17 minutes}
        \label{fig:sb_004}
    \end{subfigure}  
    \caption{Lagrangian relaxation convergence using Simulated Bifurcation for instances $t=-0.02$ and $t=0.04$, the two instances for which an optimal solution was found.}
\end{figure}

We repeated the experiment using Simulated Bifurcation as the QUBO solver with \texttt{mode='discrete'} and \texttt{heated=True}. Even with this most promising
configuration, identified testing the slack variables approach, the Lagrangian relaxation approach failed to find any feasible solution for all Markowitz instances and for the case $t=-0.005$. 
In this last, no optimal solution was identified within 200~iterations; nevertheless, a
suboptimal solution within $1.5\%$ of the optimum was obtained (Figure~\ref{fig:sb_0055}).
For $t=0.04$, an optimal solution was reached at iteration~34
(Figure~\ref{fig:sb_004}); while numerically distinct from the solution returned by \Gurobi{} (both for the QUBO and constrained formulations), it represents the same financial portfolio.
Notably, for $t=-0.02$, Simulated Bifurcation identified the optimal solution at
iteration~60 (Figure~\ref{fig:sb_002}), which \Gurobi\_Q had failed to reach.

Overall, the behaviour of the Simulated Bifurcation solver is highly unpredictable: as visible across all convergence plots, the objective value exhibits pronounced oscillations across iterations and, compared to \Gurobi\_Q, a substantially higher proportion of iterates yield infeasible solutions.

\end{document}